\documentclass[11pt]{article}
\usepackage{lineno}
\usepackage[T1]{fontenc}

\usepackage{graphicx}
\usepackage{mathptmx}      % use Times fonts if available on your TeX system
\usepackage{flushend}
\usepackage[numbers,sort&compress]{natbib}
\usepackage[colorlinks,citecolor=blue,urlcolor=blue,linkcolor=blue]{hyperref}

\usepackage{amssymb}
\usepackage{amsmath}
\usepackage{multirow}
\usepackage{bm}
\usepackage{stfloats}

\begin{document}

\centerline{\LARGE EUROPEAN ORGANIZATION FOR NUCLEAR RESEARCH}
\vspace{10mm} {\flushright{
CERN-EP-2025-012 \\
February 7, 2025\\
\vspace{4mm}
Revised version:\\
March 28, 2025\\
}}
\vspace{30mm}
\begin{center}
{\bf {\Large{Search for hadronic decays of feebly-interacting particles at NA62}}}
\end{center}

\begin{center}
{\Large The NA62 Collaboration
\footnote{Corresponding author: Jan Jerhot\\
email : jan.jerhot@cern.ch}}\\
\end{center}

\begin{abstract}
       The NA62 experiment at CERN has the capability to collect data in a beam-dump mode, where 400~GeV protons are dumped on an absorber. In this configuration, New Physics particles, including dark photons, dark scalars, and axion-like particles, may be produced in the absorber and decay in the instrumented volume beginning approximately 80~m downstream of the dump. A search for these particles decaying in flight to hadronic final states is reported, based on an analysis of a sample of $1.4 \times 10^{17}$ protons on dump collected in 2021. No evidence of a New Physics signal is observed, excluding new regions of parameter spaces of multiple models.
\end{abstract}
\vspace{15mm}
\begin{center}
    \textit{Accepted for publication in EPJC}
\end{center}

\newpage

\section{Introduction}

Fixed-target experiments present an opportunity to search for the production and decay of long-lived New Physics (NP) particles $X$ with masses, $m_X$, up to a few tens of $\mathrm{GeV}/c^2$. These experiments can operate at high intensities in a low-background environment, measuring coupling strengths, $C_X$, of NP particles to Standard Model (SM) particles in the range $10^{-8}$--$10^{-4}$. This range of $m_X$ and $C_X$ is of particular interest in models that describe hypothetical mediators between Dark Matter (DM) and SM particles, collectively referred to as dark sector portals. These mediators enable interactions between the SM and the DM sector and potentially explain various observations for which the SM does not account~\cite{Antel:2023hkf}.

A classification of dark sector portal benchmark models proposed in Ref.~\cite{Beacham:2019nyx} to facilitate the interpretation of experimental results is summarized in table \ref{tab:portals}. % Table \ref{tab:portals} summarises these models along with the SM final states that can be searched for in NP particle decays at fixed-target experiments.
The following benchmark cases (\textit{BC}s) are considered in this work:
\begin{itemize}
    \item In \textit{BC1} a new $U(1)$ symmetry gauge boson $A^\prime$, called the dark photon, interacts with the SM through kinetic mixing: $\mathcal{L} = - \varepsilon / ( 2 \cos \theta_{W})\, F^{\prime}_{\mu\nu}B^{\mu\nu}$, where $F^{\prime}_{\mu\nu}$ and $B_{\mu\nu}$ are the field strength tensors of the dark photon and the SM hypercharge gauge boson, respectively, and $\theta_{W}$ is the Weinberg angle. The strength of the interaction is characterised by $\varepsilon$, the mixing parameter with the photon.
    \item In \textit{BC4}--\textit{5} a new scalar singlet $S$, called the dark scalar, interacts with the SM Higgs doublet $H$: $\mathcal{L} = (\mu S + \lambda S^2)\, H^{\dag} H$, where $\mu$ and $\lambda$ are the coupling constants. Below the electroweak symmetry breaking scale, the dark scalar mixes with the SM Higgs boson, $h$, in proportion to the parameter $\theta \simeq \mu v / (m_h^2 - m_S^2)$, where $v$ is the vacuum expectation value of the Higgs field.
    \item In \textit{BC9}--\textit{11} an axion-like particle $a$ couples to the SM fermions $f$ and gauge bosons $V$: $\mathcal{L} = C_{ff}/(2\Lambda) \, \partial_{\mu}  a \bar{f}\gamma^{\mu}\gamma^5 f$ and $\mathcal{L} =  g^2(C_{VV}/\Lambda)\, a V_{\mu\nu}\tilde{V}^{\mu\nu}$, where $g$ is the corresponding SM gauge boson coupling constant, $\Lambda$ is the NP energy scale and $C_{ff}$ and $C_{VV}$ are the axion-like particle coupling constants. 
\end{itemize}

A search for hadronic decays of feebly-interacting particles using the NA62 beam-dump dataset corresponding to $1.4 \times 10^{17}$ protons on dump collected in 2021 is reported here. A combination of results with the previous NA62 searches for di-lepton decays~\cite{NA62:2023qyn,NA62:2023nhs} is also presented.

\begin{table*}[h]
    \def\arraystretch{1.3}
       \begin{center}

       \caption{\label{tab:portals} Summary of NP benchmark models, particle types, couplings, and decay channels relevant for fixed-target experiments assuming all possible charge combinations. Hadronic final states are highlighted; those containing exactly two oppositely charged particles are studied in this work while BC6–8 (leptonic or semi-leptonic decays) and BC9 (di-photon decay) are not considered. \vspace{1mm}}
       \begin{tabular}{|l|l|l|l|l|l|} 
       
              \hline
             benchmark \hspace{-2mm} & NP particle ($X$) &  type  & $C_X$ & \multicolumn{2}{l|}{ decay $\left(m_X < \mathcal{O}(1\,\mathrm{GeV}/c^2)\right)$}  \\ \hline\hline
              \textit{BC1} & dark photon ($A^{\prime}$) & vector & $\varepsilon$ &  $\ell\ell$ & $\boldsymbol{\pi\pi}$, $\boldsymbol{3\pi}$, $\boldsymbol{4\pi}$, $\boldsymbol{K\bar{K}}$, $\boldsymbol{K\bar{K}\pi}$ \\ \hline
             \textit{BC4--5} &  dark scalar ($S$) & scalar & $\theta$  & $\ell\ell$ & $\boldsymbol{\pi\pi}$, $\boldsymbol{4\pi}$, $\boldsymbol{K\bar{K}}$ \\ \hline
             \textit{BC9--11} & axion-like particle ($a$)  & pseudoscalar & $C_{ff,VV}$ & $\gamma\gamma$, $\ell\ell$ & $\boldsymbol{\pi\pi\gamma}$, $\boldsymbol{3\pi}$, $\boldsymbol{4\pi}$, $\boldsymbol{\pi\pi\eta}$, $\boldsymbol{K\bar{K}\pi}$  \\ \hline
             \textit{BC6--8} & heavy neutral lepton ($N_I$) \hspace{-2mm} & fermion & $U_{\alpha I}$ & \multicolumn{2}{l|}{$\pi\ell$, $\pi\pi\ell$, $K\ell$, $\ell_1\ell_2\nu$} \\ \hline
       \end{tabular}
      \vspace{-14mm}
       \end{center}
\end{table*}

% The following sections describe a search for these mediators at the NA62 experiment. The detector setup is introduced in section~\ref{sec:NA62}.
% The analysis described in section~\ref{sec:analysis} expands the physics case covered in the previous search for di-lepton decays~\cite{NA62:2023qyn,NA62:2023nhs} by including hadronic decays. The interpretation of the di-lepton analyses in various models and the combination of the di-lepton and hadronic analysis results are discussed in section \ref{sec:result}.

\section{Beamline, detector and dataset}
\label{sec:NA62}

\begin{figure}[h]
       \begin{center}
              \includegraphics[width=.55\textwidth]{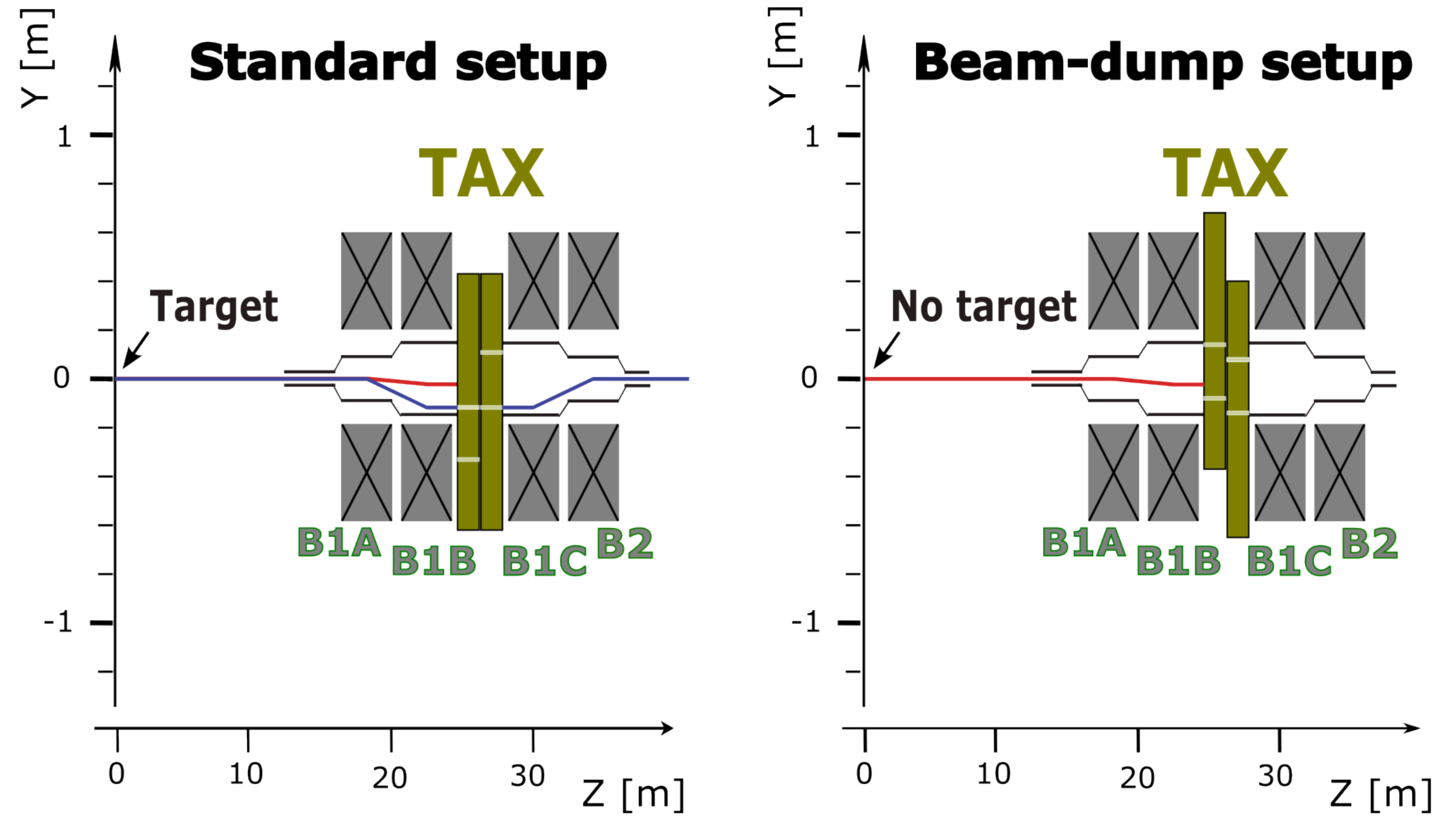}
              \vspace{-4mm}
              \caption{\label{fig:na62_tax}Schematic side view of the NA62 achromat area in the standard (left) and beam-dump (right) setups. The trajectory of a $75\,\mathrm{GeV}/c$ positively charged particle is drawn in blue while the trajectory of a $400\,\mathrm{GeV}/c$ proton is drawn in red.}
              \vspace{-10mm}
       \end{center}
\end{figure}

\begin{figure*}[h]
       \centering
       \includegraphics[width=1.0\linewidth]{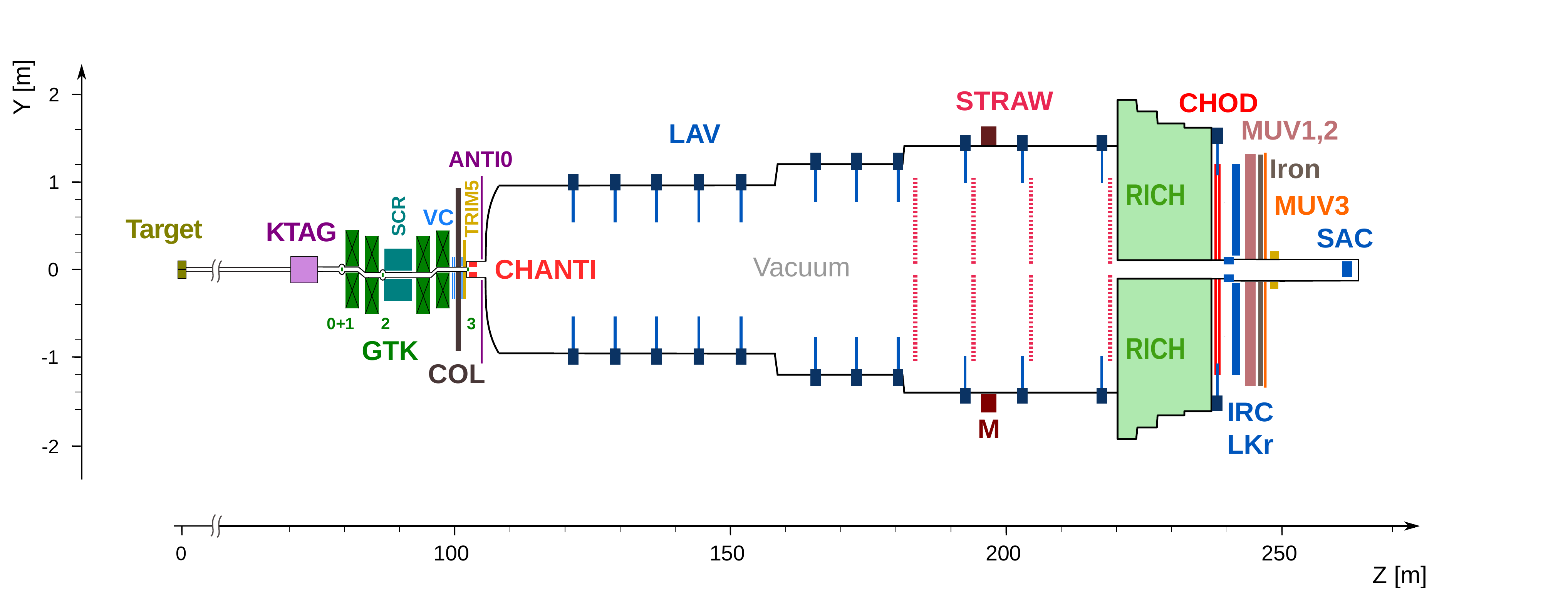}
      \vspace{-1mm}
       \caption{\label{fig:na62_layout} Schematic side view of the NA62 setup in 2021. Information from KTAG, GTK, CHANTI and VC is not used in this analysis. Not all beamline elements are shown.}
      \vspace{-1mm}
\end{figure*}

NA62 is a multi-purpose fixed-target experiment at the CERN SPS covering a broad kaon and beam-dump physics program with the main aim of measuring the ultra-rare $K^+ \to \pi^+\nu\bar\nu$ decay~\cite{NA62:2024pjp}. In the kaon operating mode, a $400\,\mathrm{GeV}/c$ proton beam from the SPS impinges on a beryllium target producing a secondary unseparated hadron beam comprising protons, pions and kaons. The position of the target defines the origin of the coordinate system.  A $75\,\mathrm{GeV}/c$ momentum component containing $6\%$ of $K^+$ is selected using an achromat formed by a set of movable copper-iron collimators called the TAX and four dipole magnets (B1A, B1B, B1C, B2), as shown in figure~\ref{fig:na62_tax}-left. In the beam-dump operating mode, the target is removed, and the TAX is moved into a position where the collimator holes do not overlap, effectively serving as a dump for the proton beam, as shown in figure~\ref{fig:na62_tax}-right. Feebly-coupled NP particles can be produced either directly in the interactions of the proton beam with the TAX, or in the interactions and decays of the secondary SM particles produced in the primary interaction. The NP particles can traverse the TAX material and reach a decay volume beginning approximately 80~m downstream. A residual flux of charged particles, most notably muons, penetrates the TAX material. To suppress the rate of these particles reaching the detector, the magnetic fields of the B1C and B2 magnets are optimised~\cite{Rosenthal:2019qua}. About $50\%$ higher proton beam intensity with respect to the kaon mode is used to maximise the NP particle production, limited by radiation protection constraints.
The beamline and detectors are schematically shown in figure~\ref{fig:na62_layout}. Further details of the beam-dump operating mode are given in Ref.~\cite{NA62:2023qyn}.

The NA62 detector~\cite{NA62:2017rwk} includes a $117\;\mathrm{m}$ long vacuum vessel beginning approximately 105\,m downstream of the target, housing a magnetic spectrometer (STRAW) to measure the momenta of charged particles. A large-aperture veto detector (ANTI0), comprising scintillator tiles, is installed upstream of the vacuum vessel~\cite{NA62:2024pjp}. The vessel is followed by a ring-imaging Cherenkov counter (RICH), used for charged-track timing and particle identification (PID), and two scintillator hodoscopes (labelled CHOD in figure \ref{fig:na62_layout}) comprising a matrix of tiles (CHOD) and two planes of slabs (NA48-CHOD) used for charged track timing.
Further PID is performed using information from the liquid krypton electromagnetic calorimeter (LKr) and iron-scintillator sandwich hadronic calorimeters (MUV1 and MUV2). Additional muon identification is provided by a muon detector (MUV3), located behind a $80\;\mathrm{cm}$-thick iron wall. The calorimeters are complemented by a small-angle veto system (SAV) and a large-angle veto system (LAV); the latter comprises 12 stations installed inside and downstream of the vacuum vessel.

Two trigger lines were implemented for the beam-dump operation in 2021: a minimum bias  trigger (Q1), requiring at least one signal in CHOD and downscaled by a factor of 20; and a two-track  trigger (H2), requiring two in-time signals in different CHOD tiles. These are complemented by a control trigger requiring at least one LKr cluster with more than $1\,\mathrm{GeV}$ deposited energy. This work describes the analysis of 2021 beam-dump data collected during 10 days of operation and corresponding to $N_{\text{POT}} = (1.4 \pm 0.3) \times 10^{17}$ protons on TAX (POT). The POT measurement is performed using a titanium-foil secondary-emission monitor located upstream of the TAX. The uncertainty of $N_\mathrm{POT}$ is deduced from the operational experience of these monitors  with independent POT measurements using activation foils and the number of selected $K^+\to\pi^+\pi^+\pi^-$ decays in the kaon operating mode, confirming a 20\% uncertainty to be a conservative estimate of the systematic error.

\section{Search for hadronic final states}
\label{sec:analysis}

\subsection{Signal selection}
\label{sub:selection}

The signal selection is applied to a sample satisfying the H2 trigger line. Exactly two good quality STRAW tracks are required; the tracks must be oppositely-charged and form a secondary vertex inside a fiducial volume (FV) with longitudinal coordinate $Z_\text{VTX}$ within $[105\,\mathrm{m}, 180\,\mathrm{m}]$ and radial transverse shape defined in Ref.~\cite{NA62:2023nhs}. The extrapolated position of each track must be in the geometrical acceptance of LKr, MUV1--3, hodoscopes and inner aperture of the last LAV station. The positions of the two tracks extrapolated to the first STRAW station and to the LKr front plane must be spatially separated by at least $20\,\mathrm{mm}$ and $200\,\mathrm{mm}$, respectively. The time of each track is defined as the time of the geometrically associated NA48-CHOD signal if present, otherwise of the CHOD signal. The time of each track must be within $5\,\mathrm{ns}$ of the trigger time. The event time is defined as the average of the two track times. No MUV3 signals are allowed within $5\,\mathrm{ns}$ and geometrically associated to either track. The two tracks must be identified as hadrons with a probability above 80\% by a boosted decision tree classifier using information from LKr, MUV1 and MUV2~\cite{NA62:2024pjp}. This condition optimises the acceptance for hadronic final states, with a 95\% efficiency per single hadron track, and the probability of mistagging a lepton as a hadron below 1\%. The RICH acceptance is optimised for the identification of positively charged particles: $K^+$ mesons can be identified with inefficiency below 5\% and $\pi^+$ meson mistagging probability as $K^+$ below 5\% in the momentum range $80$--$180\,\mathrm{GeV}/c$. Therefore, if a positive hadron is identified by the RICH as a $K^+$, the final state is classified as $K^+K^-$, otherwise as $\pi^+\pi^-$. As this selection is based only on charged particles, it includes final states containing additional photons.

To reconstruct neutral decay products, a search is performed for LKr clusters with energy exceeding $5\,\mathrm{GeV}$ within $4\,\mathrm{ns}$ from both track times and not spatially associated with any of the charged tracks. The selected clusters are assumed to belong to photons originating from the secondary vertex. If two photons are within $5\,\mathrm{ns}$ from each other, their invariant mass is used to identify $\pi^0$ or $\eta$ mesons. This enables the reconstruction of all hadronic final states from table \ref{tab:portals}. 
 
To suppress backgrounds and to avoid event misreconstruction, no in-time LAV or SAV signals are allowed. In addition, no in-time signals are allowed in ANTI0 tiles within $20\,\mathrm{cm}$ from the extrapolated tracks.

\begin{figure}[h]
       \begin{center}
              \includegraphics[width=.47\textwidth]{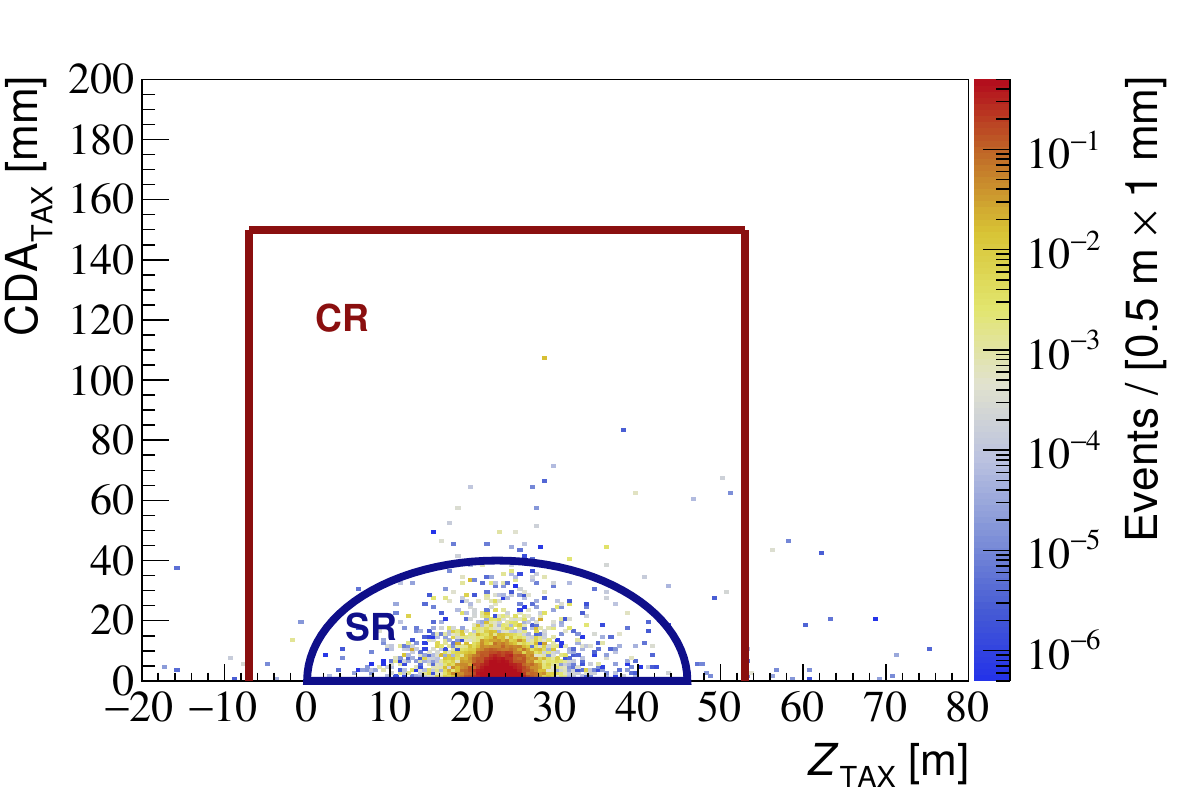}
              \hspace{5mm}
              \includegraphics[width=.47\textwidth]{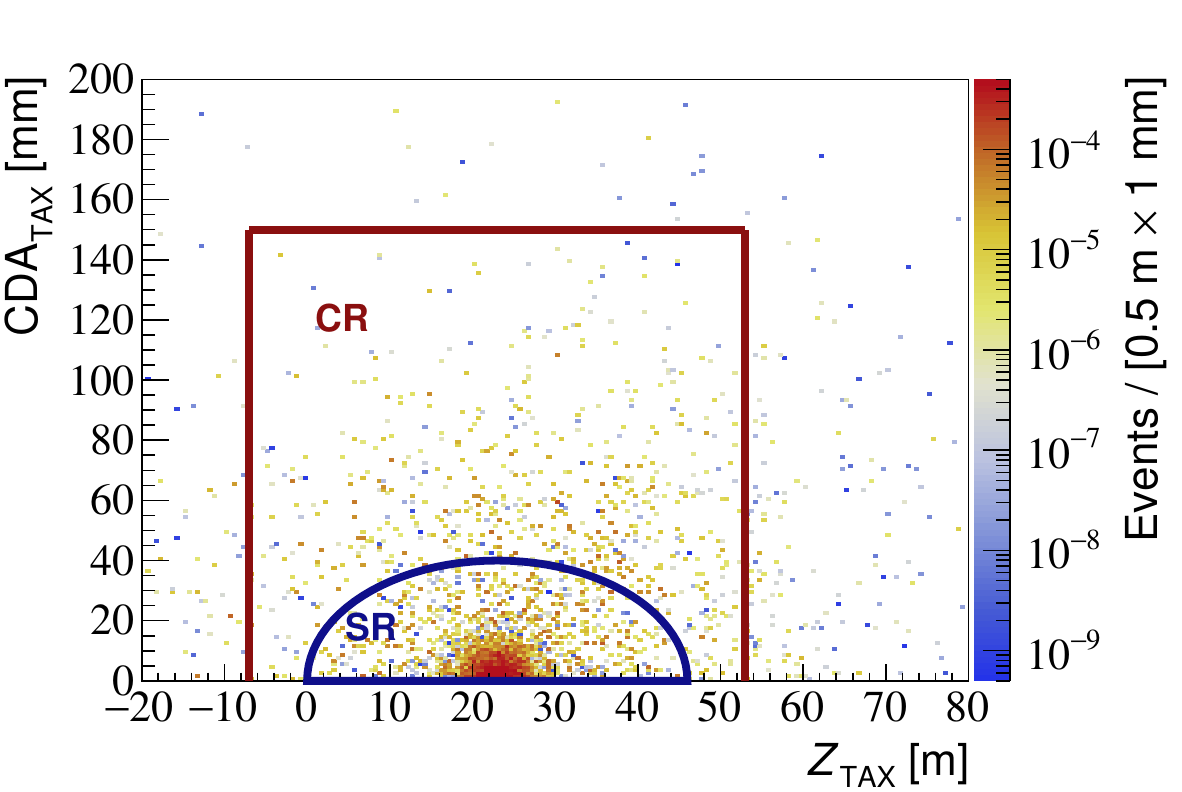}
              \vspace{-1mm}
              \caption{\label{fig:cda_vs_zvtx} Distribution of fully reconstructed simulated $A^\prime \to \pi^+ \pi^-$ (left) and $A^\prime \to \pi^+ \pi^- \pi^0\pi^0$ (right) decays in the plane  ($Z_\mathrm{TAX}$, $\mathrm{CDA}_\mathrm{TAX}$). The ellipse and box define the signal and control regions, respectively. The expected number of events is shown for \textit{BC1} model with $m_{A^\prime} = 908\,\mathrm{MeV}/c^2$, $\varepsilon = 7\times 10^{-7}$ and $N_\text{POT} = 1.4\times 10^{17}$.}
              \vspace{-5mm}
       \end{center}
\end{figure}

The three-momentum of the NP particle candidate is calculated as the sum of the three-momenta of the reconstructed decay products and is used to extrapolate the particle trajectory backward from the secondary vertex to the closest approach to the primary proton beam axis. The mid-point of the minimum-distance segment between the two lines defines the primary vertex where the NP particle is produced. 
The distributions of simulated $A^\prime \to \pi^+\pi^-$ and $A^\prime \to \pi^+\pi^-\pi^0\pi^0$ decays in the plane ($Z_\mathrm{TAX}$, $\mathrm{CDA}_\mathrm{TAX}$) are shown in figure~\ref{fig:cda_vs_zvtx}, where $Z_\mathrm{TAX}$ is the $Z$ coordinate of the primary vertex and $\text{CDA}_\text{TAX}$ is the closest distance of approach between the NP particle trajectory and the beam axis. The other NP decays considered in table~\ref{tab:portals} show similar properties. The signal region (SR) is the area inside a half-ellipse centred at (23\,m, 0\,mm), corresponding to the mean proton beam impact point on TAX, with semi-axes 23\,m and 40\,mm, respectively. The control region (CR) is a rectangle surrounding the SR defined as $-7\,\mathrm{m} < Z_\text{TAX} < 53\,\mathrm{m}$ and $\text{CDA}_\text{TAX}<150\,\mathrm{mm}$. Both regions are kept masked in the data sample until the background estimates are validated. 

The mass of the candidate NP particle $m_X$ is computed using the masses and three-momenta of the reconstructed particles at the secondary vertex. For the signal, a Gaussian distribution of $m_X$ is expected, with a standard deviation $\sigma_{m_X}$ varying with the NP particle mass. 

\subsection{Background estimation}
\label{sub:background}

\begin{figure}[!t]
       \begin{center}
              \includegraphics[width=0.49\textwidth]{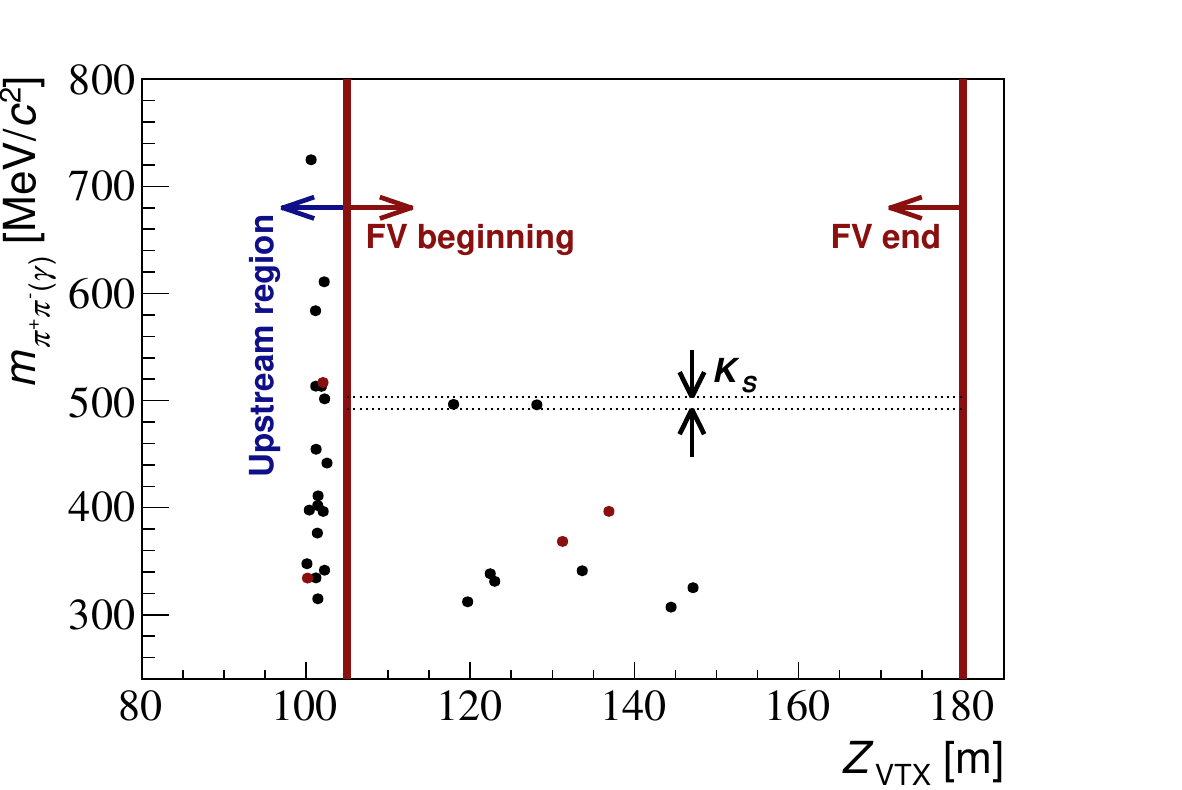}
              \caption{\label{fig:bkg_invmass}Distributions of $\pi^+\pi^-$ (black) and $\pi^+\pi^-\gamma$ (red) data events in the plane ($ Z_\mathrm{VTX}$, $m_{\pi\pi(\gamma)}$) when inverting the ANTI0 veto condition and removing the LAV veto condition. Vertical solid lines indicate the FV. The excluded $3\sigma$ window around the $K_S$ mass is indicated by horizontal dashed lines.}
              \vspace{-5mm}
       \end{center}
\end{figure}

\begin{figure}[!b]
       \begin{center}
              \includegraphics[width=0.49\textwidth]{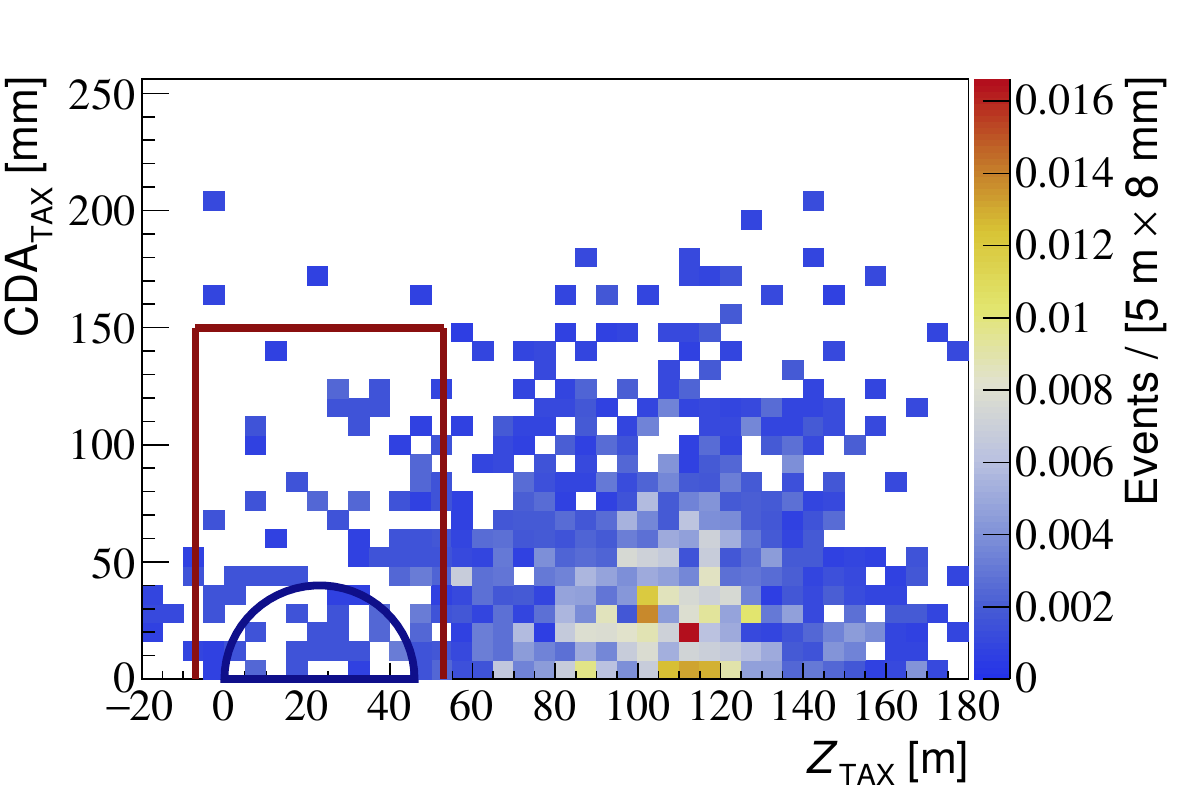}
              \caption{\label{fig:bkg_yield}Distribution of simulated $K^+ \to \pi^+\pi^+\pi^-$ decay events in the plane ($Z_\mathrm{TAX}$, $\mathrm{CDA}_\mathrm{TAX}$)        without the LAV and ANTI0 veto conditions, scaled to the observed number of data events.}
              \vspace{-5mm}
       \end{center}
\end{figure}

The background sources are studied and their estimates are validated outside the masked SR and CR using simulations and data-driven methods. 
With the number of expected $\ell^+\ell^-$ background events at the level of $10^{-2}$~\cite{NA62:2023qyn,NA62:2023nhs} and the probability to misidentify a charged lepton as a hadron below $10^{-2}$, only direct hadron production can lead to sizeable backgrounds. Four types of processes resulting in hadron production are identified: 
\begin{itemize}
       \item \textit{Kaon decays:} Hadrons escaping the TAX can interact upstream of the FV and produce secondary particles including kaons. Figure~\ref{fig:bkg_invmass} shows the distribution of reconstructed $\pi^+\pi^-(\gamma)$ data events in the plane ($ Z_\mathrm{VTX}$, $m_{\pi\pi(\gamma)}$) when inverting the ANTI0 veto condition and removing the LAV veto condition. The distribution consists of three components: interactions in the collimator preceding the FV; $K_S\to\pi^+\pi^-$ decays; $K^+\to \pi^+\pi^+\pi^-$ decays with a pion escaping detection or mis-reconstructed as a photon. The upstream interactions do not enter the signal sample as they are not reconstructed in the FV. A 3$\sigma$ window in the NP mass around the $K_S$ mass is ignored for all final states. To evaluate the $K^+$ background, single tracks, collected by the Q1 trigger and identified as $K^+$ using the RICH, are used as an input for the $K^+ \to \pi^+\pi^+\pi^-$ decay simulations in the FV. Oversampling of the decay in the FV by a factor of 100 enhances the statistical power of the simulation, allowing an estimate of the background contamination in the SR and CR after applying the full selection. The resulting $\pi^+\pi^-$ and $\pi^+\pi^-\gamma$ mass distributions are empirically fitted using  Gaussian functions with mean values of $340\,\mathrm{MeV}/c^2$ and $450\,\mathrm{MeV}/c^2$, and standard deviations of $10\,\mathrm{MeV}/c^2$ and $40\,\mathrm{MeV}/c^2$, respectively. The simulated distribution of the background events in the ($Z_\mathrm{TAX}$, $\mathrm{CDA}_\mathrm{TAX}$) plane obtained without applying LAV and ANTI0 veto conditions is shown in figure~\ref{fig:bkg_yield}.  
       
       \item \textit{Prompt:} The prompt background produced by muons traversing the material  upstream of or within the vacuum vessel is evaluated with a data-driven Monte Carlo (MC) simulation. Single tracks, collected by the Q1 trigger and identified as muons by LKr and MUV3, are used as an input for a standalone code (\texttt{PUMAS}~\cite{Niess:2017rlv}) interfaced with \texttt{GEANT4}~\cite{GEANT4:2002zbu}. Muons are propagated backwards accounting for the expected energy loss and bending induced by magnetic fields. The  muon energy and geometrical distribution obtained in a plane upstream of the FV is used as an input for a forward \texttt{GEANT4}-based simulation of the muon interactions in the detector material. The background mechanism identified with the simulation is inelastic muon-nucleus interaction. The prompt background is found to contribute to each final state but at the level of $10^{-4}$ events or less. 

       \item \textit{Combinatorial:} The combinatorial background originates from the pairing of interaction products of uncorrelated beam protons. This contribution is evaluated using single tracks collected by the Q1 trigger identified as hadrons, overlaid in time to simulate accidental superposition. While this component is responsible for the dominant background for the $\mu\mu$ analysis~\cite{NA62:2023qyn}, with approximately 0.02 hadron tracks per muon track it results in a $\pi^+\pi^-$ background of the order of $10^{-5}$ events, and therefore it is negligible for the hadronic final states.
       
       \item  \textit{Neutrino-induced:} The flux of $\nu_\mu$ and $\overline{\nu}_\mu$ corresponding to $10^{18}$ proton interactions in the TAX is determined using \texttt{GEANT4}. Charged and neutral current interactions are simulated in the passive material using the \texttt{GENIE} framework~\cite{Andreopoulos:2009rq}, effectively enhancing the interaction cross section, interfaced with \texttt{GEANT4}. No two-track events are reconstructed. The background is found to be negligible.
   \end{itemize}

The background estimates are summarized in table~\ref{tab:bkg}. 
The $K^+ \to \pi^+\pi^+\pi^-$ decay constitutes the dominant background process and contributes to the $\pi^+\pi^-$ and $\pi^+\pi^-\gamma$\, final states only.

\begin{table*}[!h]

           \begin{center}
           \caption{\label{tab:bkg}Summary of expected background counts $N_\text{exp}$ in CR and SR after full selection, with errors corresponding to a $68\%$ coverage, and the minimum number of events $N_{\text{min}}^{5\sigma}$ to be observed to claim a $5\sigma$ discovery, separately for SR and the union of SR and CR.\vspace{1mm}}
                     
           \begin{tabular}{|l|c|c|c|c|} 
              \hline
              Channel & $N_{\text{exp,CR}}$ & $N_{\text{exp,SR}}$ & $N_{\text{min,SR}}^{5\sigma}$ & $N_{\text{min,SR+CR}}^{5\sigma}$ \\
              \hline\hline
              $\pi^+\pi^-$ & $0.013 \pm 0.007$ & $0.007 \pm 0.005$ & 3 & 4\\
              $\pi^+\pi^-\gamma$ & $0.031 \pm 0.016$ & $0.007 \pm 0.004$ & 3 & 5 \\
              $\pi^+\pi^-\pi^0$   &  $(1.3^{+4.4}_{-1.0})\times10^{-7}$ &  $(1.2^{+4.3}_{-1.0})\times10^{-7}$ & 1 & 1 \\
              $\pi^+\pi^-\pi^0\pi^0$ & $(1.6^{+7.6}_{-1.4})\times10^{-8}$ &  $(1.6^{+7.4}_{-1.4})\times10^{-8}$ & 1 & 1 \\
              $\pi^+\pi^-\eta$    &  $(7.3^{+27.0}_{-6.1})\times10^{-8}$ &  $(7.0^{+26.2}_{-5.8})\times10^{-8}$ & 1 & 1 \\
              $K^+K^-$            &  $(4.7^{+15.7}_{-3.9})\times10^{-7}$ &  $(4.6^{+15.2}_{-3.8})\times10^{-7}$ & 1 & 2 \\
              $K^+K^-\pi^0$       &  $(1.6^{+3.2}_{-1.2})\times10^{-9}$ &  $(1.5^{+3.1}_{-1.2})\times10^{-9}$ & 1 & 1 \\
              \hline
          \end{tabular}
          \end{center}
          \vspace{-10mm}
\end{table*}

\subsection{Expected number of signal events}
\label{sub:signal}
The numbers of expected signal events $N_\mathrm{exp}^{ij}$ as a function of the particle mass $m_X$ and lifetime $\tau_X$ are estimated for each combination of production process $i$ and final state $j$ using \texttt{GEANT4}-based simulations. To allow a model-independent interpretation of the analysis result, the coupling $C_X$ is kept at a reference value, considered as a multiplicative constant, and the decay branching ratio $\mathrm{BR}^j_X$ is assumed to be unity for each channel $j$. The hadronic decay channels $j$ are highlighted in table \ref{tab:portals} for each NP particle.  Including the di-lepton decay channels studied in Ref.~\cite{NA62:2023nhs}, 61 combinations of production processes and decay channels are studied. The values of $N_\mathrm{exp}^{ij}$ are evaluated as

\begin{equation}
       % \begin{split}
       \label{eq:nexp}
              N_\text{exp}^{ij}(m_X,\tau_X) = N_{\text{POT}} \times \chi^i_{pp \to X}(m_X) \times P^i_\text{RD}(m_X,\tau_X) \times A^{ij}_\text{acc}(m_X,\tau_X)\;,
       % \end{split}
\end{equation}
where the probability for the NP particle to reach the FV and decay therein, $P^i_\text{RD}$, and the signal selection acceptance, $A_\mathrm{acc}^{ij}$, for NP particles that reach the FV and decay therein, including trigger efficiency, are functions of $m_X$ and $\tau_X$. The probability of NP particle production in the dump, $\chi^i_{pp \to X}$, is evaluated for several production processes:
\begin{itemize}
       \item \textit{$B$-meson decays for dark scalars and axion-like particles:} $B$ mesons are produced by interactions of the primary beam protons in the TAX. The $B^+$, $B^-$, $B^0$ and $\bar{B}^0$ production kinematic spectra are simulated with \texttt{PYTHIA8.3}~\cite{Bierlich:2022pfr} under the conservative assumption that $B$ mesons are produced only by primary interactions of the beam protons with the TAX. The production cross section derived in Ref.~\cite{Schubert:2024hpm} is used. 
       The $B \to K X$ decays, where $K$ stands for both charged and neutral kaons and their resonances, are simulated, using the decay widths from Refs.~\cite{Bauer:2021mvw} and~\cite{Boiarska:2019jym} for the model-dependent interpretation for axion-like particles and dark scalars, respectively.
      
       \item \textit{Light-meson decays for dark photons:} Light pseudoscalar $P = \lbrace \pi^0$, $\eta$, $\eta^\prime \rbrace$ and  vector $V = \lbrace \rho^0$, $\omega$, $\phi \rbrace$ mesons are produced by interactions of the primary beam protons in the TAX. The light meson spectra are simulated using \texttt{PYTHIA8.3}, validated in Ref.~\cite{Dobrich:2019dxc}. The $P \to A^\prime \gamma$ and $V \to A^\prime P$ decays are considered.
       
       \item \textit{Meson mixing for dark photons and axion-like particles:} 
       The light mesons produced as in the previous production process can mix with the NP particles carrying the same quantum numbers. The kinematics of the emitted NP particles is modified with respect to the light mesons by $P-a$ and $V-A^\prime$ mixing, but there is no complete treatment of this effect available in the literature.
       Therefore, the NP particle kinematics is approximately evaluated as discussed in Ref.~\cite{Jerhot:2022chi}.
       
       \item \textit{Bremsstrahlung production for dark scalars and dark photons:} NP particles can be produced through quasi-elastic scattering of the beam proton on the TAX nuclei. The dark scalar brems-strahlung is simulated using the quasi-real approximation from Ref.~\cite{Foroughi-Abari:2021zbm}. 
       A time-like form factor accounts for the resonant production enhancement. An off-shell form factor effectively decreases the particle production when the energy transfer approaches a cutoff scale of $1.5\,\mathrm{GeV}$. 
       The dark photon bremsstrahlung is simulated using the modified Weizsacker-Williams approximation as in Refs.~\cite{NA62:2023qyn,NA62:2023nhs}. For comparison, results using this approximation, not accounting for the off-shell form factor but including the time-like form factor are also derived. 
       \item \textit{Primakoff production for axion-like particles:} Axion-like particles can be resonantly produced through the Primakoff effect. This can be mediated by both off-shell photons from primary protons and on-shell photons from light meson decays. The differential cross section from Ref.~\cite{Dobrich:2015jyk} is used.
\end{itemize}

\begin{figure}[!b]
       \begin{center}
              \vspace{-3mm}
              \includegraphics[width=0.49\textwidth]{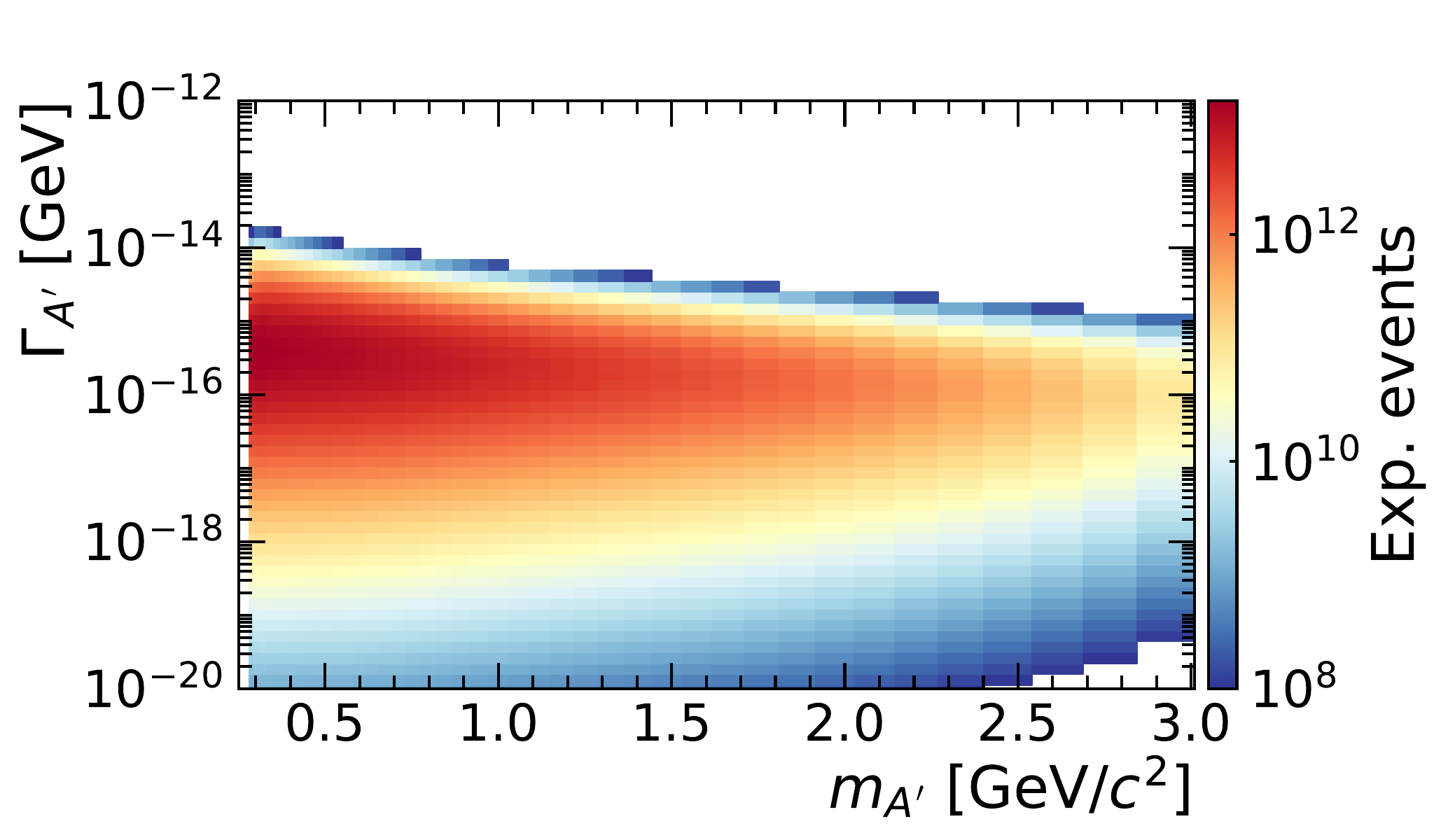}
              \includegraphics[width=0.49\textwidth]{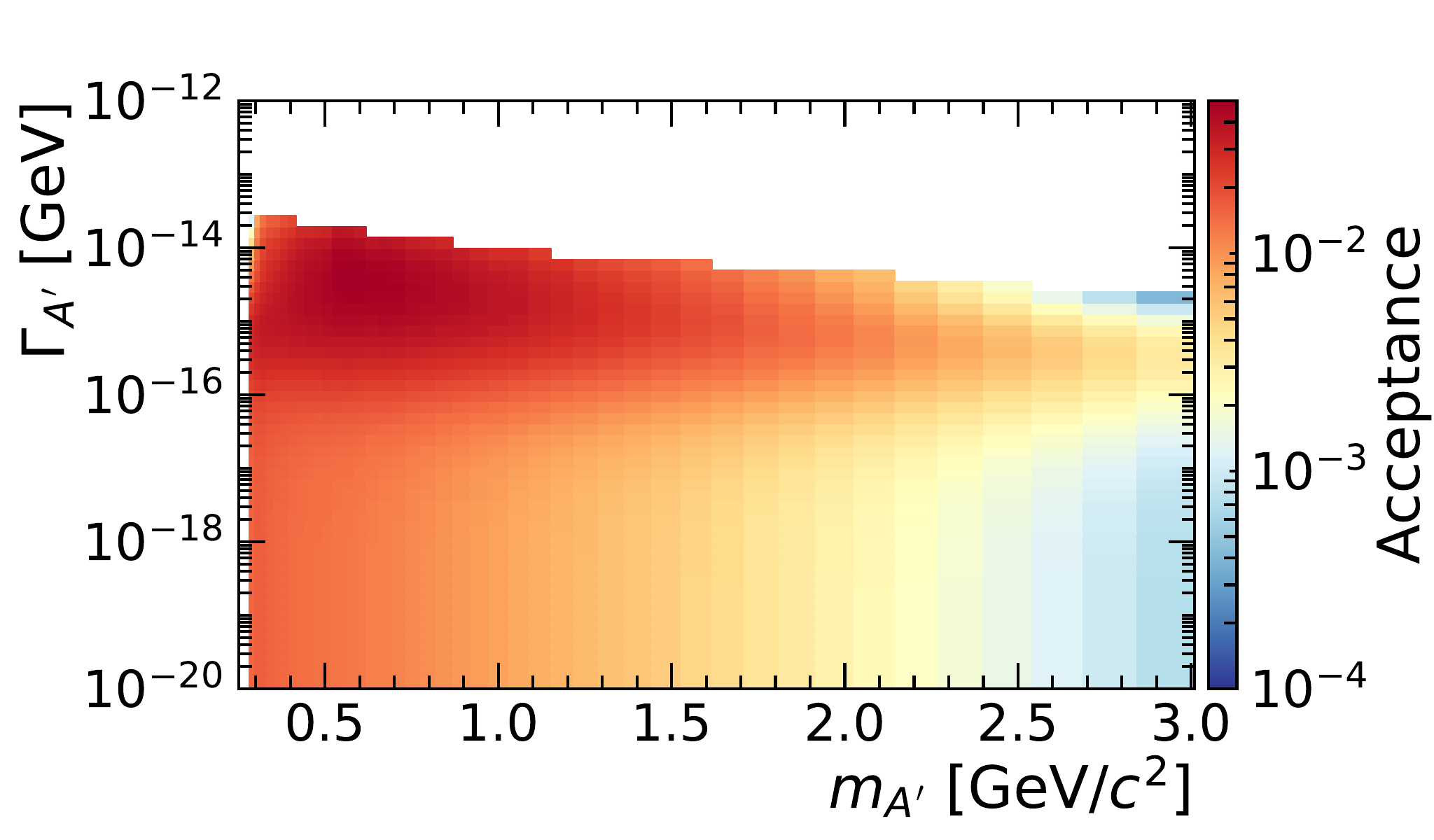}\\
              \vspace{-1mm}
              \includegraphics[width=0.49\textwidth]{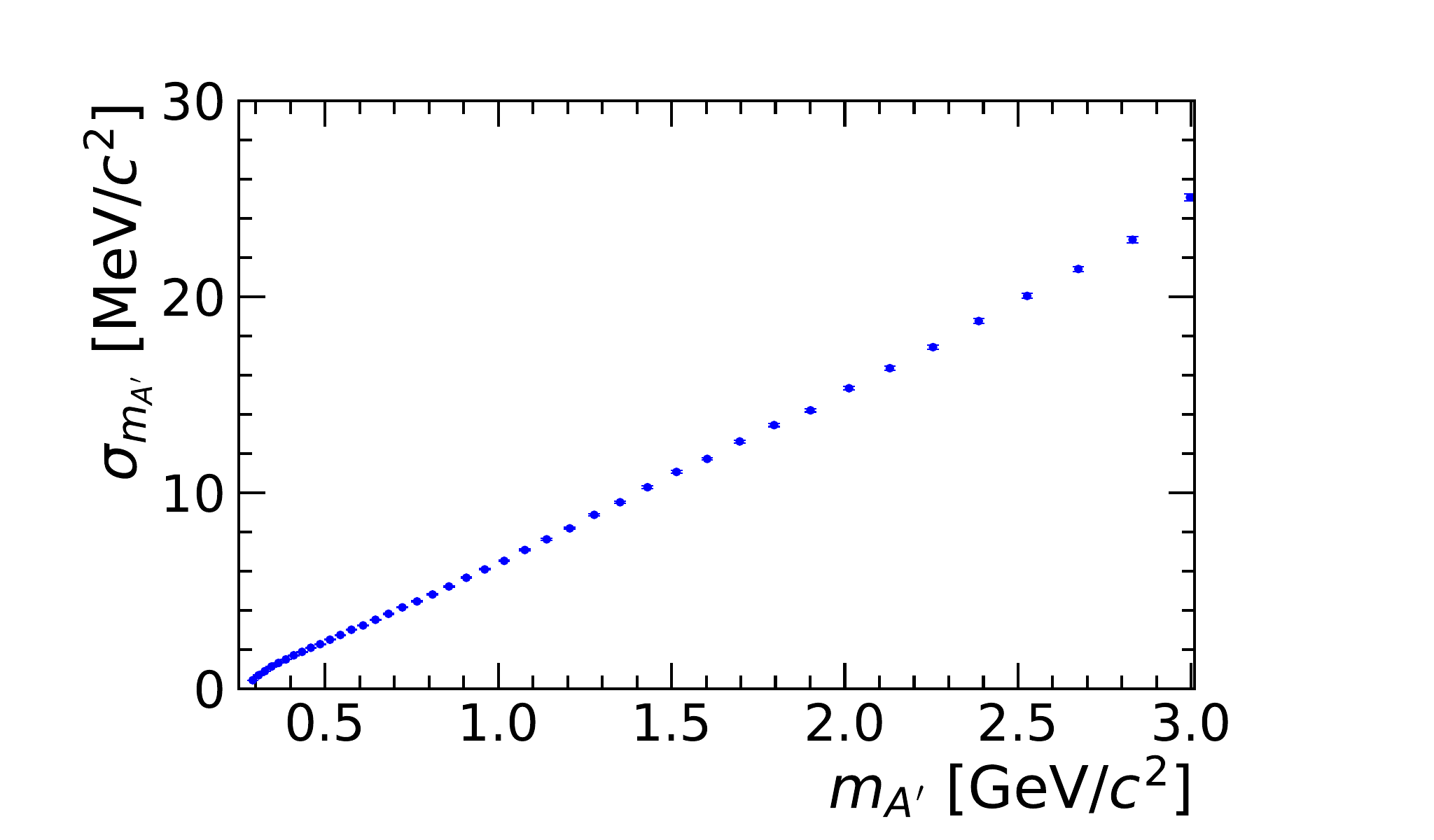}~\hspace{-5mm}
              \vspace{-3mm}
              \caption{\label{fig:mc_acceptance}Top left: Number of expected $A^\prime \to \pi^+\pi^-$ events in the plane ($m_{A^\prime}$, $ \Gamma_{A^\prime} = \hslash / \tau_{A^\prime}$) after full selection, assuming $\varepsilon = 1$ and $\mathrm{BR}(A^\prime \to \pi^+ \pi^-) = 1$. Top right: Acceptance of the $A^\prime \to \pi^+ \pi^-$ decay for particles that reach the FV and decay therein  shown in the plane ($m_{A^\prime}$, $ \Gamma_{A^\prime}$). Bottom: Mass resolution of the reconstructed NP particle in the decay $A^\prime \to \pi^+ \pi^-$ as a function of its mass.}
       \end{center}
       \vspace{-5mm}
\end{figure}

  Figure \ref{fig:mc_acceptance} summarises the information obtained using equation~\ref{eq:nexp} and mass resolution for a dark photon produced via bremsstrahlung and decaying to a $\pi^+\pi^-$ pair.

  The systematic uncertainty in the expected number of events, computed from equation (\ref{eq:nexp}), is dominated by the 20\% relative uncertainty in the measured $N_\mathrm{POT}$. The theory uncertainties entering $P_{RD}$ and $\chi_{pp \to X}$ are not considered. The systematic uncertainty in $A_\text{acc}$ is estimated to be $3.3\%$, dominated by the limited simulation statistics and the simulation of PID efficiencies.

\section{Results}
\label{sec:result}
After unmasking the CRs and SRs, no events are observed for the hadronic decays studied. The public framework \texttt{ALPINIST}~\cite{Jerhot:2022chi}
is used for the model-dependent interpretation of the result for each \textit{BC} scenario, calculating the total expected number of events as a function of the $X$ mass and coupling, $N_\text{exp}(m_X,C_X)$, by combining the individual values of $N^{ij}_\text{exp}(m_X,\tau_X)$ according to table~\ref{tab:portals} including the di-lepton channels studied in Ref.~\cite{NA62:2023nhs}. 
The decay widths of the hadronic channels are calculated from Refs.~\cite{Ilten:2018crw,Aloni:2018vki,Winkler:2018qyg,DallaValleGarcia:2023xhh}.

The exclusion limits are derived using the $\text{CL}_s$ method~\cite{Junk:1999kv} performing a likelihood fit to the observation on a grid of $C_X$ and $m_X$ values using the profile likelihood ratio from Ref.~\cite{NA62:2023nhs} as a test statistic. The $N_\text{exp}(m_X,C_X)$ distributions obtained for each decay channel and the expected number of background events from table~\ref{tab:bkg} enter as nuisance parameters in the likelihood function. The $N_\mathrm{POT}$ and the background estimate probability distribution functions are modelled as log-normal distributions. The expected signal and background mass distributions described in sections~\ref{sub:selection} and~\ref{sub:background} are also included in the likelihood evaluation.

\begin{figure}[!b]
       \begin{center}
              \vspace{-3mm}
              \includegraphics[width=0.48\textwidth]{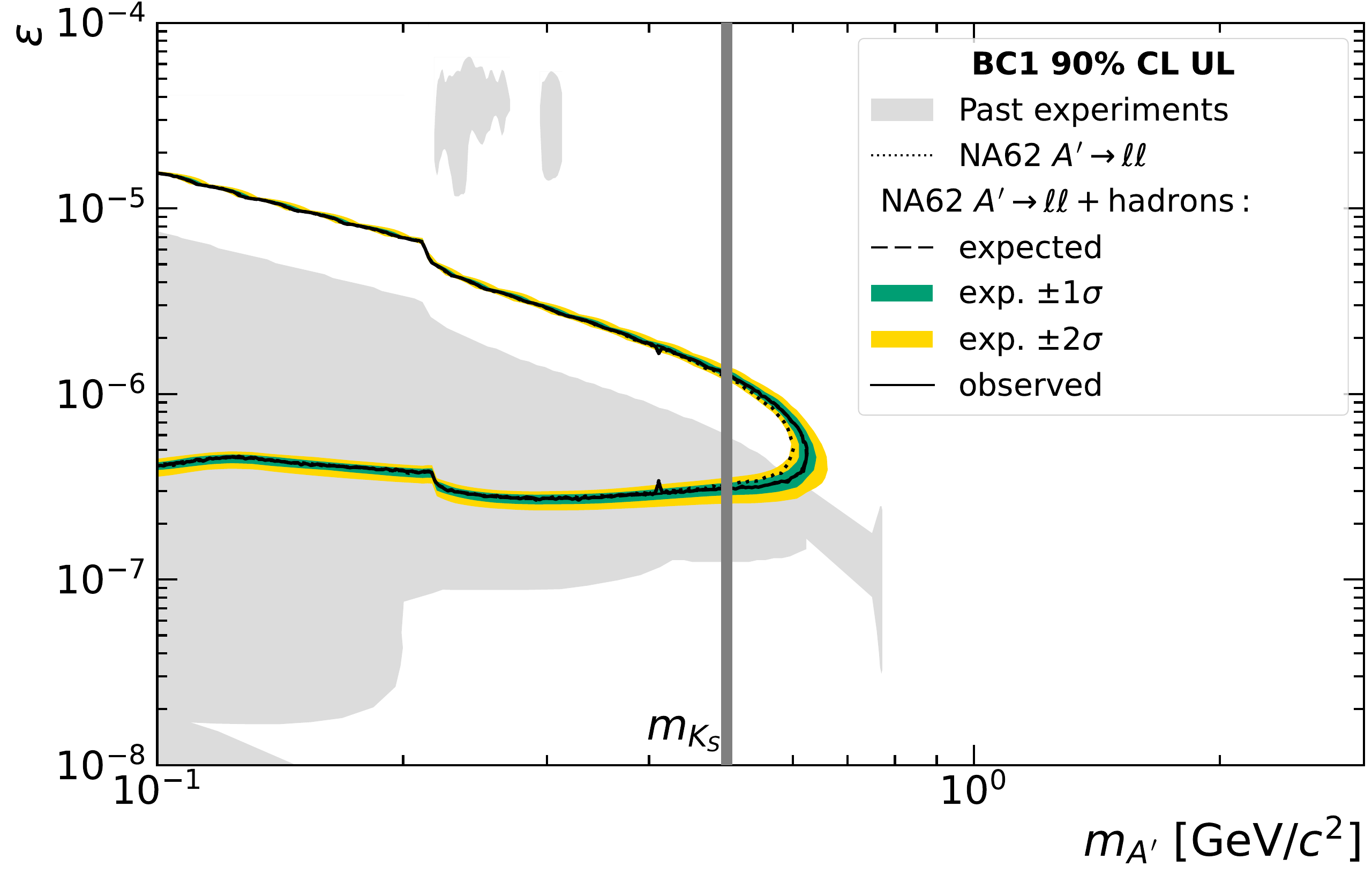}
              \includegraphics[width=0.48\textwidth]{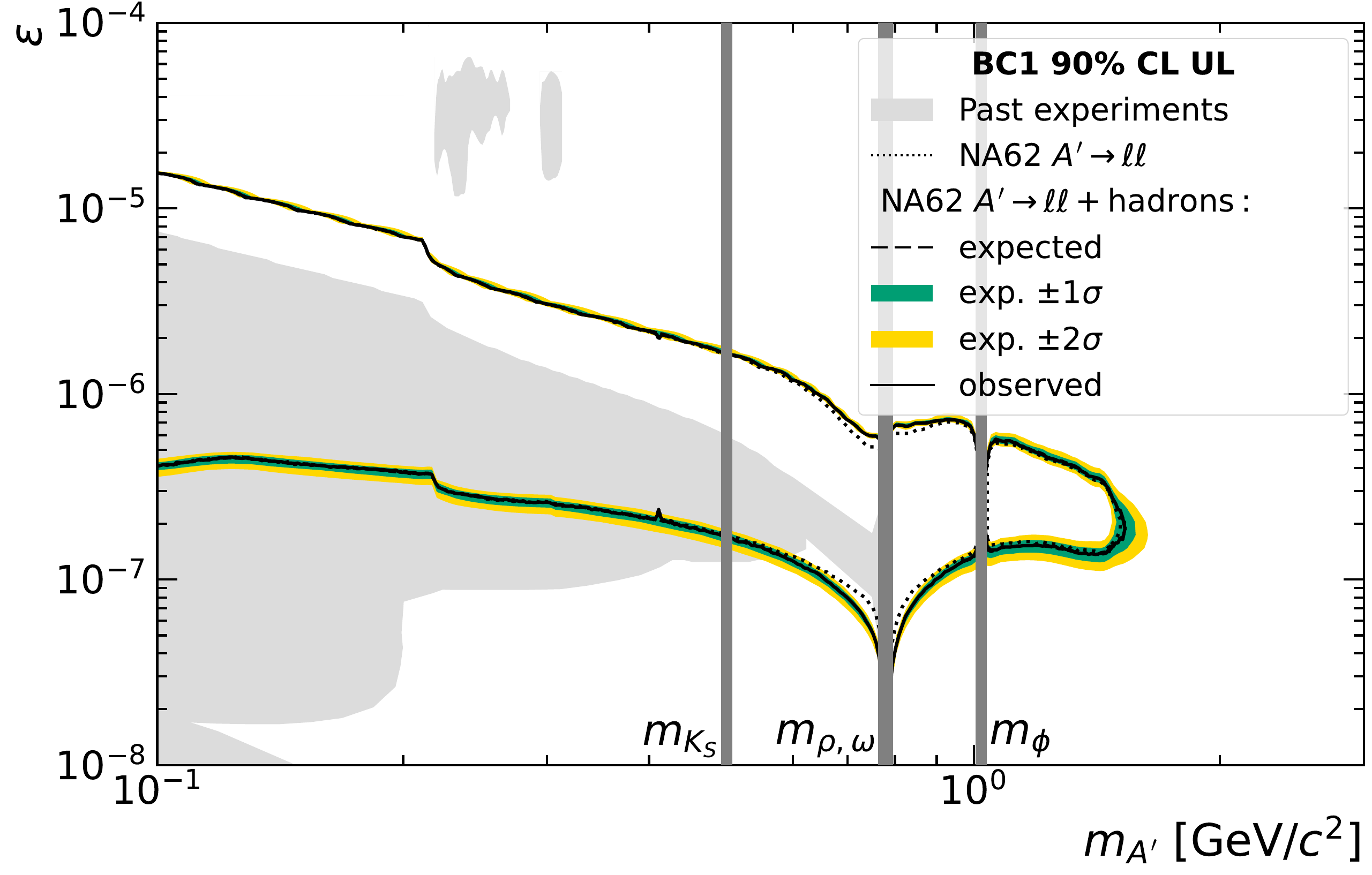}
              \vspace{-3mm}
              \caption{\label{fig:exclusion_dp}
              Observed 90\% CL exclusion contours in the plane ($m_{A^\prime}$, $\varepsilon$) in dark photon \textit{BC1} benchmark case combining hadronic and di-lepton channels compared to the updated NA62 di-lepton result. Left: Result using bremsstrahlung production without the time-like form factor. Right: Result including mixing production and bremsstrahlung production with a time-like form factor. Expected $\pm1\sigma$ and $\pm2\sigma$ bands correspond to the uncertainty in the number of protons on TAX (theory uncertainty not included).
              In both panels, the exclusion contours for past proton beam-dump experiments assume a bremsstrahlung production including a time-like form factor.}
       \end{center}
       \vspace{-9mm}
\end{figure}

\begin{figure}[!h]
       \begin{center}
              \includegraphics[width=0.50\textwidth]{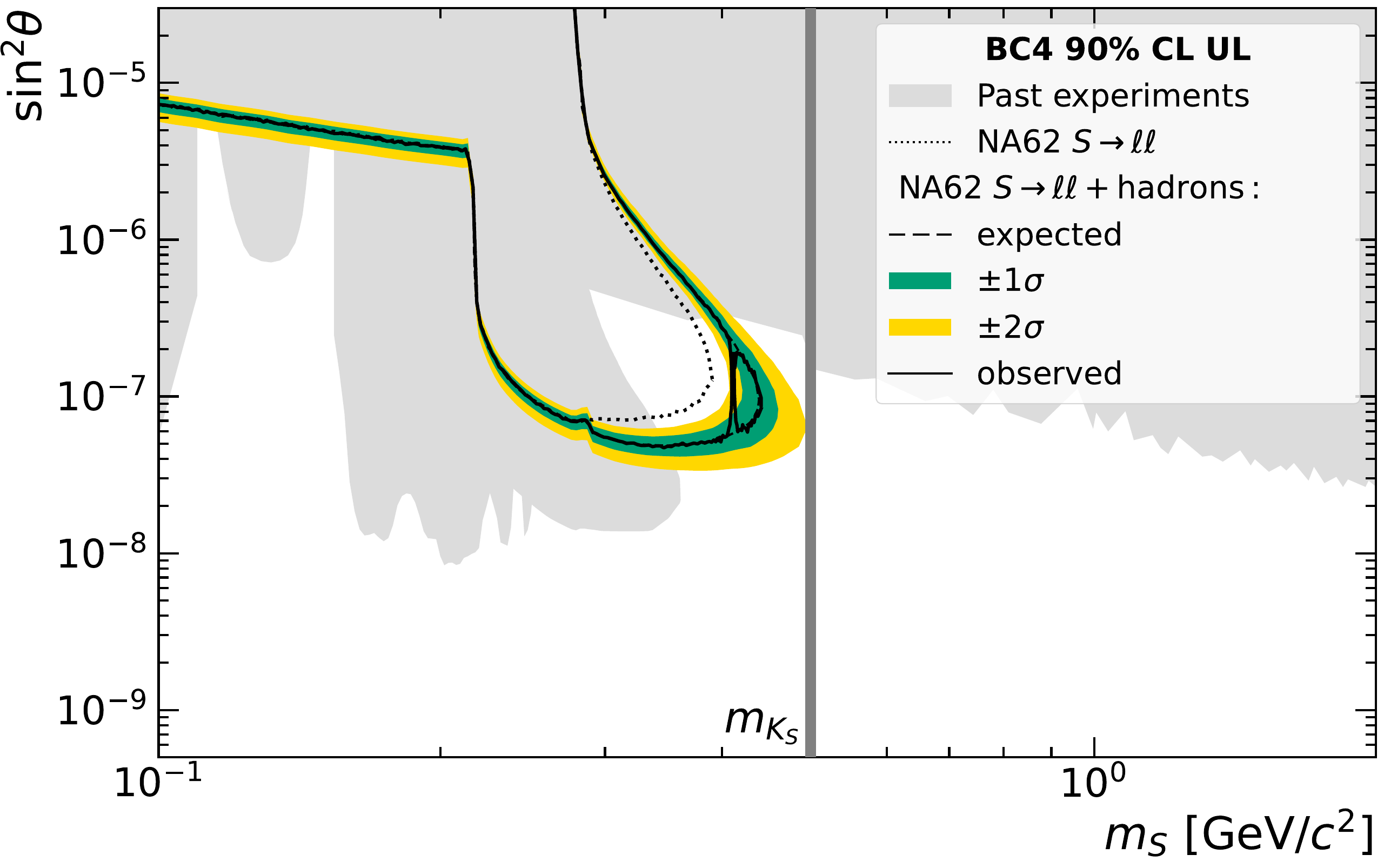}
       
              \vspace{-3mm}
              \caption{\label{fig:exclusion_ds}
              The observed 90\% CL exclusion contours in the plane ($m_{S}$, $\mathrm{sin}^2\theta$)  in dark scalar \textit{BC4} benchmark case combining hadronic and di-lepton channels compared to the NA62 di-lepton result. Expected $\pm1\sigma$ and $\pm2\sigma$ bands correspond to the uncertainty in the number of protons on TAX (theory uncertainty not included).
              }
       \end{center}
       \vspace{-8mm}
\end{figure}

\begin{figure}[!h]
       \begin{center}
              \includegraphics[width=0.48\textwidth]{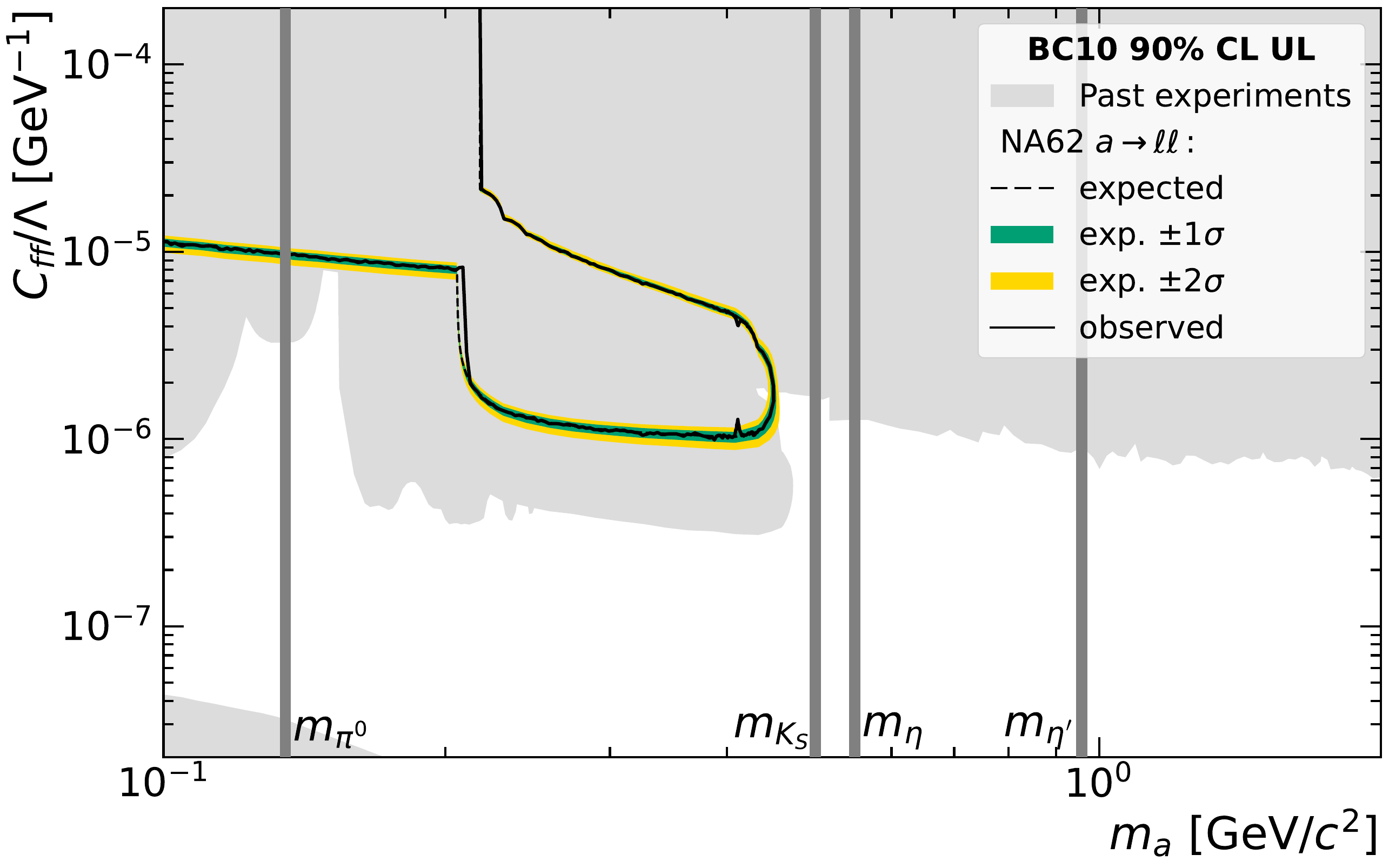}
              \includegraphics[width=0.505\textwidth]{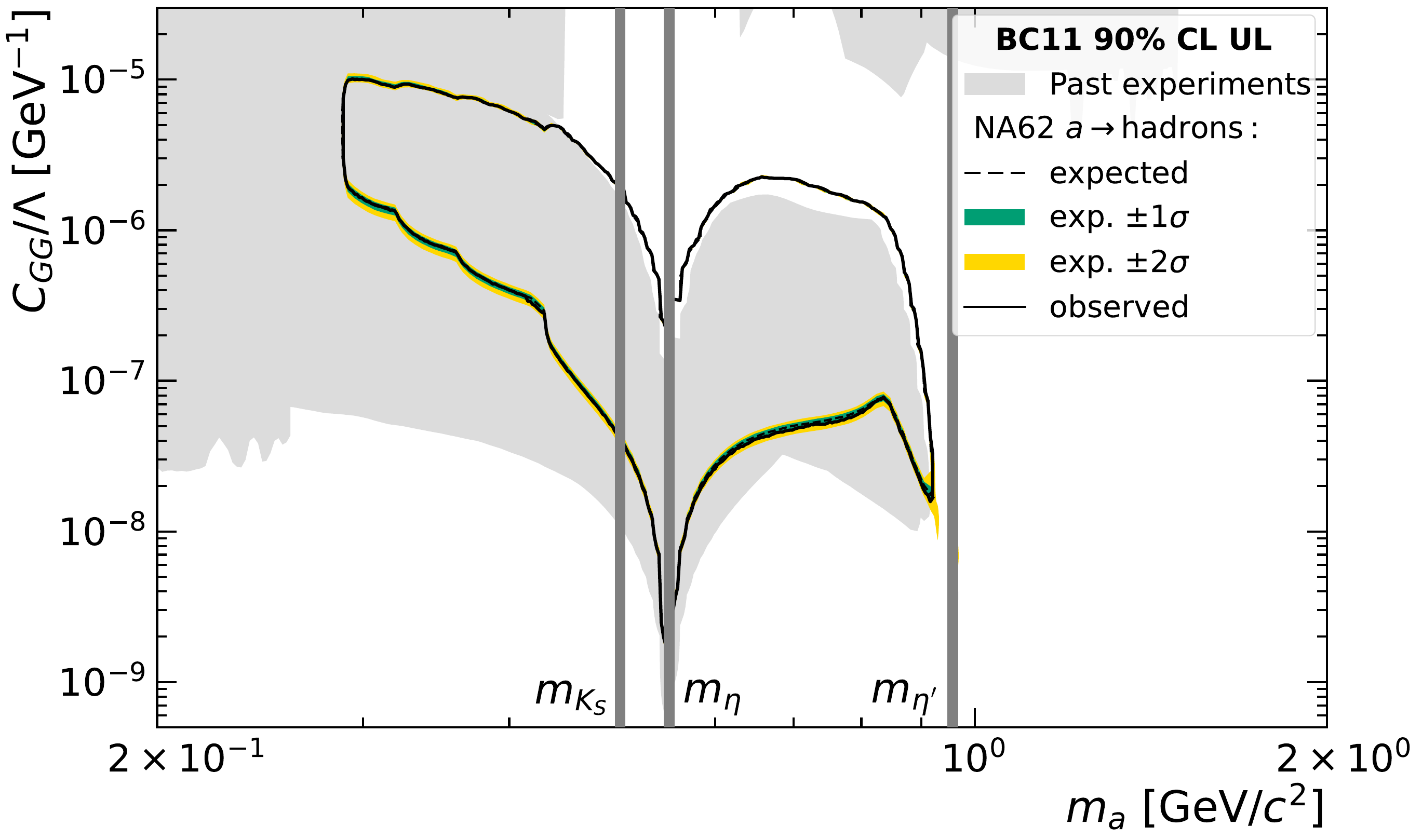}
              \vspace{-3mm}
              \caption{\label{fig:exclusion_alp} 
              The observed 90\% CL exclusion contours in the plane ($m_{a}$, $C_a/\Lambda$) in the fermion-coupled axion-like particle \textit{BC10} (left) and gluon-coupled axion-like particle \textit{BC11} (right) benchmark cases, evaluated assuming $\Lambda = 1\,\mathrm{TeV}$. Expected $\pm1\sigma$ and $\pm2\sigma$ bands correspond to the uncertainty in the number of protons on TAX (theory uncertainty not included).
              }
       \end{center}
       \vspace{-6mm}
\end{figure}

\begin{figure}[!h]
       \begin{center}
              \includegraphics[width=0.49\textwidth]{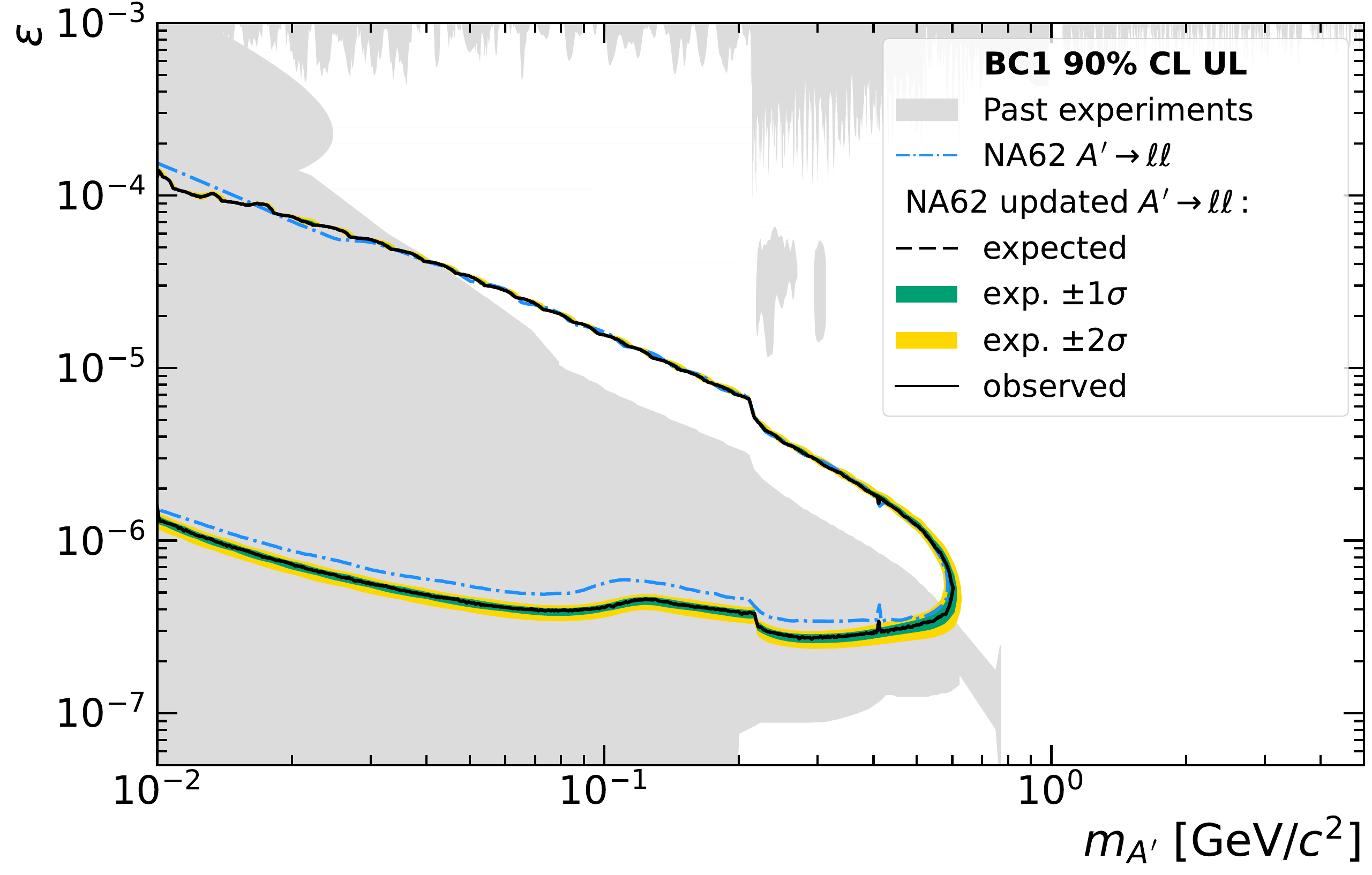}
              \includegraphics[width=0.49\textwidth]{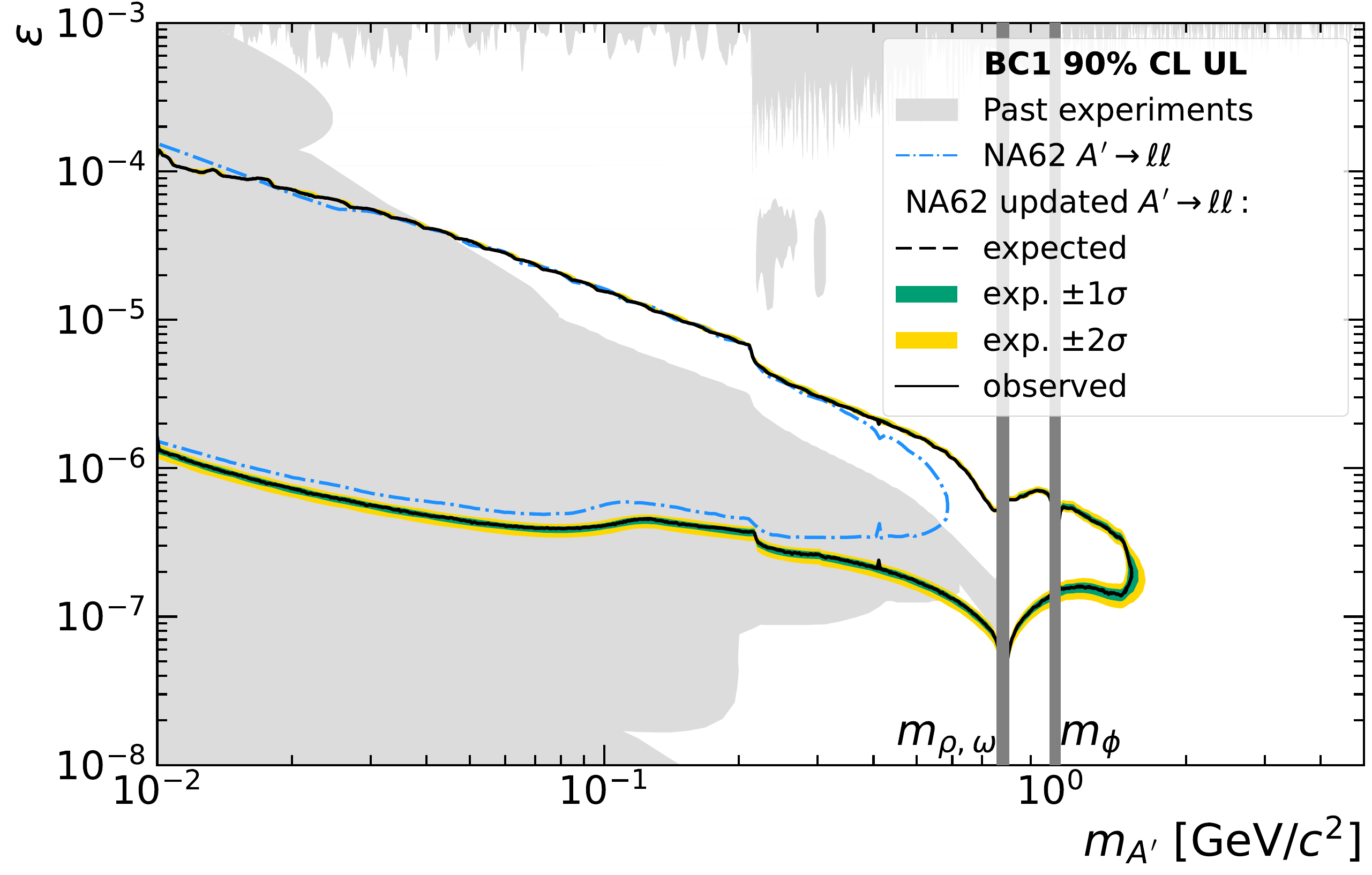}
              \vspace{-3mm}
              \caption{\label{fig:dp_dilepton}
              The observed 90\% CL exclusion contours in the plane ($m_{A^\prime}$, $\varepsilon$) in \textit{BC1} benchmark case for di-lepton final states together with the expected $\pm1\sigma$ and $\pm2\sigma$ bands (theory uncertainty not included) with updated light meson spectra only (left) and with mixing production and time-like form factor for bremsstrahlung production (right). The exclusion contour obtained in~\cite{NA62:2023nhs} is displayed as a dash-dotted blue line.}
       \end{center}
       \vspace{-8mm}
\end{figure}

The results for \textit{BC1} (dark photon) and \textit{BC4} (dark scalar) are shown in figures \ref{fig:exclusion_dp} and \ref{fig:exclusion_ds}, respectively. The result for \textit{BC5} is equivalent to \textit{BC4} as the exclusion limit does not extend to the low coupling region, in which the dark scalar pair production in $B$ meson decays becomes relevant. The axion-like particle exclusion bounds shown in figure~\ref{fig:exclusion_alp} are evaluated assuming a UV scale $\Lambda = 1\,\mathrm{TeV}$. Due to small hadronic decay widths for \textit{BC10} (fermion-coupled axion-like particle), only the di-lepton decays are considered. Similarly for \textit{BC11} (gluon-coupled axion-like particle), only hadronic decay channels are considered. Mass windows around $\pi^0, \eta, \eta^\prime, \rho, \omega, \phi$ masses are not displayed in figures~\ref{fig:exclusion_dp}-\ref{fig:exclusion_alp} as the theory estimate of the expected signal is not reliable for NP particles quasi-degenerate with respect to the SM particles. The expected di-lepton signal counts are updated with respect to the evaluation used in Ref.~\cite{NA62:2023nhs} by extending to the full momentum range of the light mesons. For completeness, a comparison between the updated di-lepton exclusion bound with the one published in Ref.~\cite{NA62:2023nhs} is shown in figure \ref{fig:dp_dilepton}. 

In all studied benchmark cases the exclusion contours extend beyond the previous limits: the proton beam-dump experiments CHARM and NuCal~\cite{CHARM:1985anb,Blumlein:1990ay,Blumlein:1991xh,Blumlein:2013cua,Jerhot:2022chi}, the electron beam-dump experiments E137, E141 and NA64~\cite{Bjorken:1988as,Riordan:1987aw,Andreas:2012mt,NA64:2020qwq,NA64:2019auh}, the forward collider experiment FASER~\cite{FASER:2023tle,FASER:2024bbl} and the kaon-decay measurements by NA62 and E949~\cite{NA62:2021zjw,Afik:2023mhj,NA62:2020pwi,NA62:2023olg,E949:2005qiy,BNL-E949:2009dza,Gori:2020xvq}. 

\section{Conclusions and prospects}
The NA62 2021 beam-dump data sample equivalent to $N_\text{POT} = 1.4 \times 10^{17}$ has been investigated for decays of New Physics particles into $\pi^+\pi^-$, $\pi^+\pi^-\gamma$, $\pi^+\pi^-\pi^0$, $\pi^+\pi^-\pi^0\pi^0$, $\pi^+\pi^-\eta$, $K^+K^-$ and $K^+K^-\pi^0$ final states, with no signal observed. Combining this result with the previous searches for the di-lepton final states, $e^+ e^-$ and $\mu^+ \mu^-$, using the same dataset, new regions of dark scalar, dark photon and axion-like particle parameter spaces are excluded, improving on previous experimental searches. An additional NA62 beam-dump dataset equivalent to $N_\text{POT} = 4.9 \times 10^{17}$ is being analysed. The analysis has demonstrated no background limitation for statistics of $N_{\text{POT}} = 10^{18}$.

\section*{Acknowledgements}
It is a pleasure to express our appreciation to the staff of the CERN laboratory and the technical
staff of the participating laboratories and universities for their efforts in the operation of the
experiment and data processing.

The cost of the experiment and its auxiliary systems was supported by the funding agencies of 
the Collaboration Institutes. We are particularly indebted to: 
F.R.S.-FNRS (Fonds de la Recherche Scientifique - FNRS), under Grants No. 4.4512.10, 1.B.258.20, Belgium;
CECI (Consortium des Equipements de Calcul Intensif), funded by the Fonds de la Recherche Scientifique de Belgique (F.R.S.-FNRS) under Grant No. 2.5020.11 and by the Walloon Region, Belgium;
%BMES (Ministry of Education, Youth and Science), Bulgaria;
NSERC (Natural Sciences and Engineering Research Council), funding SAPPJ-2018-0017,  Canada;
%NRC (National Research Council) contribution to TRIUMF, Canada;
MEYS (Ministry of Education, Youth and Sports) funding LM 2018104, Czech Republic;
BMBF (Bundesministerium f\"{u}r Bildung und Forschung), Germany;
% contracts 05H12UM5, 05H15UMCNA, 05H18UMCNA, 05H21UMCNA, and 05H24UMCNA numbers not requested
INFN  (Istituto Nazionale di Fisica Nucleare),  Italy;
MIUR (Ministero dell'Istruzione, dell'Uni-versit\`a e della Ricerca),  Italy;
CONACyT  (Consejo Nacional de Ciencia y Tecnolog\'{i}a),  Mexico;
IFA (Institute of Atomic Physics) Romanian 
% ended in 2019  CERN-RO No. 1/16.03.2016 
% 2020 and 2021
CERN-RO Nr. 10/10.03.2020
% 2022-2024  CERN-RO Nr. 06/03.01.2022
and Nucleus Programme PN 19 06 01 04,  Romania;
% remove for opt C
%INR-RAS (Institute for Nuclear Research of the Russian Academy of Sciences), Moscow, Russia; 
%JINR (Joint Institute for Nuclear Research), Dubna, Russia; 
%NRC (National Research Center)  ``Kurchatov Institute'' and MESRF (Ministry of Education and Science of the Russian Federation), Russia; 
MESRS  (Ministry of Education, Science, Research and Sport), Slovakia; 
CERN (European Organization for Nuclear Research), Switzerland; 
STFC (Science and Technology Facilities Council), United Kingdom;
NSF (National Science Foundation) Award Numbers 1506088 and 1806430,  U.S.A.;
ERC (European Research Council)  ``UniversaLepto'' advanced grant 268062, ``KaonLepton'' starting grant 336581, Europe.

Individuals have received support from:
% updated 2024
%Charles University (Research Center UNCE/SCI/013, grant PRIMUS 23/SCI/025), Czech Republic;
Charles University (grants UNCE 24/SCI/016, PRIMUS 23/SCI/025), Ministry of Education, Youth and Sports (project FORTE CZ.02.01.01/00/22-008/0004632), Czech Republic;
Czech Science Foundation (grant 23-06770S);
Agence Nationale de la Recherche (grant ANR-19-CE31-0009), France;
Ministero dell'Istruzione, dell'Universit\`a e della Ricerca (MIUR  ``Futuro in ricerca 2012''  grant RBFR12JF2Z, Project GAP), Italy;
% terminated  Russian Foundation for Basic Research  (RFBR grants 18-32-00072, 18-32-00245), Russia; 
% remove for opt C  Russian Science Foundation (RSF 19-72-10096), Russia;
the Royal Society  (grants UF100308, UF0758946), United Kingdom;
STFC (Rutherford fellowships ST/J00412X/1, ST/M005798/1), United Kingdom;
ERC (grants 268062,  336581 and  starting grant 802836 ``AxScale'');
EU Horizon 2020 (Marie Sk\l{}odowska-Curie grants 701386, 754496, 842407, 893101, 101023808).

\bibliography{bibliography}{}

\bibliographystyle{spphys}
\newpage

\newcommand{\orcimg}{\raisebox{-0.3\height}{\includegraphics[height=\fontcharht\font`A]{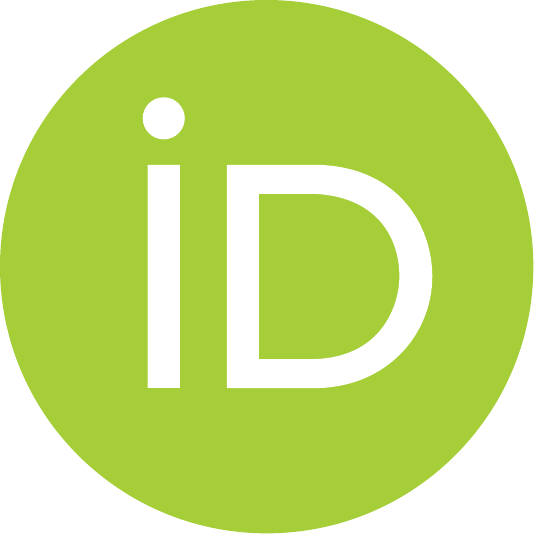}}}
\newcommand{\orcid}[1]{\href{https://orcid.org/#1}{\orcimg}}

\centerline{\bf The NA62 Collaboration}
\vspace{1cm}
%
%%%%%%%%%%%%%%%%%%%%%%%%%%%%%%%%%
%

\begin{raggedright}
\noindent
%%%%%%%
{\bf Universit\'e Catholique de Louvain, Louvain-La-Neuve, Belgium}\\
 E.~Cortina Gil\orcid{0000-0001-9627-699X},
 J.~Jerhot$\,$\renewcommand{\thefootnote}{\fnsymbol{footnote}}\footnotemark[1]\renewcommand{\thefootnote}{\arabic{footnote}}$^,$$\,${\footnotemark[1]}\orcid{0000-0002-3236-1471},
 N.~Lurkin\orcid{0000-0002-9440-5927}
\vspace{0.5cm}

%%%%%%%
{\bf TRIUMF, Vancouver, British Columbia, Canada}\\
 T.~Numao\orcid{0000-0001-5232-6190},
 B.~Velghe\orcid{0000-0002-0797-8381},
 V. W. S.~Wong\orcid{0000-0001-5975-8164}
\vspace{0.5cm}

%%%%%%%
{\bf University of British Columbia, Vancouver, British Columbia, Canada}\\
 D.~Bryman$\,${\footnotemark[2]}\orcid{0000-0002-9691-0775}
\vspace{0.5cm}

%%%%%%%
{\bf Charles University, Prague, Czech Republic}\\
 Z.~Hives\orcid{0000-0002-5025-993X},
 T.~Husek$\,${\footnotemark[3]}\orcid{0000-0002-7208-9150},
 K.~Kampf\orcid{0000-0003-1096-667X},
 M.~Kolesar\orcid{0000-0002-9085-2252},
 M.~Koval\orcid{0000-0002-6027-317X}
\vspace{0.5cm}

%%%%%%%
{\bf Aix Marseille University, CNRS/IN2P3, CPPM, Marseille, France}\\
 B.~De Martino\orcid{0000-0003-2028-9326},
 M.~Perrin-Terrin\orcid{0000-0002-3568-1956},
 L.~Petit$\,${\footnotemark[4]}\orcid{0009-0000-8079-9710}
\vspace{0.5cm}

%%%%%%%
{\bf Max-Planck-Institut f\"ur Physik (Werner-Heisenberg-Institut), Garching, Germany}\\
 B.~D\"obrich\orcid{0000-0002-6008-8601},
 S.~Lezki\orcid{0000-0002-6909-774X},
 J.~Schubert$\,${\footnotemark[5]}\orcid{0000-0002-5782-8816}
% ...
\vspace{0.5cm}

%%%%%%%
{\bf School of Physics, University of Bristol, Bristol, United Kingdom}\\
 H.~Heath\orcid{0000-0001-6576-9740}
\vspace{0.5cm}

%%%%%%%
{\bf School of Physics and Astronomy, University of Glasgow, Glasgow, United Kingdom}\\
 D.~Britton\orcid{0000-0001-9998-4342},
 A.~Norton\orcid{0000-0001-5959-5879},
 D.~Protopopescu\orcid{0000-0002-8047-6513}
\vspace{0.5cm}

%%%%%%%
{\bf Physics Department, University of Lancaster, Lancaster, United Kingdom}\\
 J. B.~Dainton,
 L.~Gatignon\orcid{0000-0001-6439-2945},
 R. W. L.~Jones\orcid{0000-0002-6427-3513}
\vspace{0.5cm}

%%%%%%%
{\bf Physics and Astronomy Department, George Mason University, Fairfax, Virginia, USA}\\
 P.~Cooper,
 D.~Coward$\,${\footnotemark[24]}\orcid{0000-0001-7588-1779},
 P.~Rubin\orcid{0000-0001-6678-4985}
\vspace{0.5cm}

%%%%%%%
{\bf Authors affiliated with an international laboratory covered by a cooperation agreement with CERN}\\
 A.~Baeva,
 D.~Baigarashev$\,${\footnotemark[25]}\orcid{0000-0001-6101-317X},
 V.~Bautin\orcid{0000-0002-5283-6059},
 D.~Emelyanov,
 T.~Enik\orcid{0000-0002-2761-9730}, 
 V.~Falaleev$\,${\footnotemark[18]}\orcid{0000-0003-3150-2196},
 V.~Kekelidze\orcid{0000-0001-8122-5065},
 D.~Kereibay,
 A.~Korotkova,
 L.~Litov$\,${\footnotemark[9]}\orcid{0000-0002-8511-6883}, 
 D.~Madigozhin\orcid{0000-0001-8524-3455},
 M.~Misheva$\,${\footnotemark[26]},
 N.~Molokanova,
 I.~Polenkevich,
 Yu.~Potrebenikov\orcid{0000-0003-1437-4129}, 
 K.~Salamatin\orcid{0000-0001-6287-8685},
 S.~Shkarovskiy
\vspace{0.5cm}

%%%%%%%
{\bf Authors affiliated with an Institute formerly covered by a cooperation agreement with CERN}\\
 S.~Fedotov,
 K.~Gorshanov\orcid{0000-0001-7912-5962},
 E.~Gushchin\orcid{0000-0001-8857-1665},
 S.~Kholodenko$\,${\footnotemark[27]}\orcid{0000-0002-0260-6570},
 A.~Khotyantsev, 
 Y.~Kudenko\orcid{0000-0003-3204-9426},
 V.~Kurochka,
 V.~Kurshetsov\orcid{0000-0003-0174-7336},
 A.~Mefodev,
 V.~Obraztsov\orcid{0000-0002-0994-3641}, 
 A.~Okhotnikov\orcid{0000-0003-1404-3522},
 A.~Sadovskiy\orcid{0000-0002-4448-6845},
 V.~Sugonyaev\orcid{0000-0003-4449-9993},
 O.~Yushchenko\orcid{0000-0003-4236-5115}
\vspace{0.5cm}

\end{raggedright}

%
%%%%%%%%%%%%%%%%%%%%%%%%%%%%%%%%%
%

\setcounter{footnote}{0}
\newlength{\basefootnotesep}
\setlength{\basefootnotesep}{\footnotesep}

\renewcommand{\thefootnote}{\fnsymbol{footnote}}
\noindent
$^{\footnotemark[1]}${Corresponding author: J.~Jerhot, email: jan.jerhot@cern.ch}\\
%$^{\footnotemark[2]}${Deceased}\\
\renewcommand{\thefootnote}{\arabic{footnote}}
$^{1}${Present address: Max-Planck-Institut f\"ur Physik (Werner-Heisenberg-Institut), D-85748 Garching, Germany} \\
$^{2}${Also at TRIUMF, Vancouver, British Columbia, V6T 2A3, Canada} \\
$^{3}${Also at School of Physics and Astronomy, University of Birmingham, Birmingham, B15 2TT, UK} \\
$^{4}${Also at Universit\'e de Toulon, Aix Marseille University, CNRS, IM2NP, F-83957, La Garde, France} \\
$^{5}${Also at Department of Physics, Technical University of Munich, M\"unchen, D-80333, Germany} \\
$^{6}${Present address: Institut f\"ur Kernphysik and Helmholtz Institute Mainz, Universit\"at Mainz, Mainz, D-55099, Germany} \\
$^{7}${Also at CERN, European Organization for Nuclear Research, CH-1211 Geneva 23, Switzerland} \\
$^{8}${Also at Dipartimento di Scienze Fisiche, Informatiche e Matematiche, Universit\`a di Modena e Reggio Emilia, I-41125 Modena, Italy} \\
$^{9}${Also at Faculty of Physics, University of Sofia, BG-1164 Sofia, Bulgaria} \\
$^{10}${Present address: Scuola Superiore Meridionale e INFN, Sezione di Napoli, I-80138 Napoli, Italy} \\
$^{11}${Also at Department of Industrial Engineering, University of Roma Tor Vergata, I-00173 Roma, Italy} \\
$^{12}${Also at Department of Electronic Engineering, University of Roma Tor Vergata, I-00173 Roma, Italy} \\
$^{13}${Also at Universit\`a degli Studi del Piemonte Orientale, I-13100 Vercelli, Italy} \\
$^{14}${Present address: Universit\'e Catholique de Louvain, B-1348 Louvain-La-Neuve, Belgium} \\
$^{15}${Also at Gran Sasso Science Institute, I-67100 L'Aquila,  Italy} \\
$^{16}${Also at al-Farabi Kazakh National University, 050040 Almaty, Kazakhstan} \\
$^{17}${Also at Universidad de Guanajuato, 36000 Guanajuato, Mexico} \\
$^{18}${Present address: INFN, Sezione di Perugia, I-06100 Perugia, Italy} \\
$^{19}${Present address: Charles University, 116 36 Prague 1, Czech Republic} \\
$^{20}${Present address: Physics Department, University of Lancaster, Lancaster, LA1 4YB, UK} \\
$^{21}${Present address: Syracuse University, Syracuse, NY 13244, USA} \\
$^{22}${Present address: Laboratori Nazionali di Frascati, I-00044 Frascati, Italy} \\
$^{23}${Present address: Dipartimento di Fisica dell'Universit\`a e INFN, Sezione di Genova, I-16146 Genova, Italy} \\
$^{24}${Also at SLAC National Accelerator Laboratory, Stanford University, Menlo Park, CA 94025, USA} \\
$^{25}${Also at L. N. Gumilyov Eurasian National University, 010000 Nur-Sultan, Kazakhstan} \\
$^{26}${Present address: Institute of Nuclear Research and Nuclear Energy of Bulgarian Academy of Science (INRNE-BAS), BG-1784 Sofia, Bulgaria} \\
$^{27}${Present address: INFN, Sezione di Pisa, I-56100 Pisa, Italy} \\

\end{document}